\documentclass[11pt,a4paper]{article}
\usepackage{amsmath}
\usepackage[mathcal]{eucal}
\usepackage{latexsym}
\usepackage{amssymb}
\begin{titlepage}
\title{Instanton representation of Plebanski gravity. Gravitational instantons from the classical formalism}
\author{Eyo Eyo Ita III}
\medskip
\input amssym.def
\input amssym.tex

\def \in{\indent}

\begin{document}
\maketitle
\bigskip
\centerline{Department of Applied Mathematics and Theoretical Physics} 
\smallskip
\centerline{Centre for Mathematical Sciences, University of Cambridge, Wilberforce Road}
\smallskip
\centerline{Cambridge CB3 0WA, United Kingdom}
\smallskip
\centerline{eei20@cam.ac.uk} 
\bigskip  
                 
\begin{abstract}  
We present a reformulation of general relativity as a `generalized' Yang--Mills theory of gravity, using a SO(3,C) gauge connection and the self-dual Weyl tensor as dynamical variables.  This formulation uses Plebanski's theory as the starting point, and obtains a new action called the instanton representation (IRPG).  This paper has yielded a collection of various new results, which show that the IRPG is indeed a fundamentally new approach to GR which should not be confused with existing approaches.  Additionally, the IRPG appears to provide a realization of the relation amongst general relativity, Yang--Mills theory and instantons.
\end{abstract}
\end{titlepage}
 
\section{Introduction}
  
In the 1980's there was a major development in general relativity (GR) due to Abhay Ashtekar, which provided a new set of Yang--Mills like variables known as the Ashtekar 
variables (See e.g. \cite{ASH1}, \cite{ASH2} and \cite{ASH3}).  These variables have re-invigorated the efforts at achieving a quantum theory of gravity using techniques from Yang--Mills theory.  Additionally, the relation of general relativity to Yang--Mills theory by its own right is an interesting and active area of research \cite{CHAKRA}, \cite{CAP}.  The purpose of the present paper is two-fold.  First, we will provide a new formulation of GR which shows that its relation to Yang--Mills theory can be taken more literally in a certain well-defined context.  The degrees of freedom of GR will be explicitly embedded in a Yang--Mills like action resembling an instanton, and this formulation will be named the instanton representation of Plebanski gravity.  In this paper we will focus just on the classical aspects of the theory, and make contact with existing results of GR as well as provide various new results.\par
\indent
The organization of this paper is as follows.  In section 2 we will first provide a review of Plebanski's theory of gravity $I_{Pleb}$ and the mechanism by which the Einstein equations follow from it.  The Plebanski action contains a self-dual connection one-form $A^a$, where $a=1,2,3$ denotes an $SO(3,C)$ index with respect to which the (internal) self-duality is defined, a matrix $\psi_{ae}\in{SO}(3,C)\otimes{SO}(3,C)$, and a triple of self-dual 
two-forms $\Sigma^a$, also self-dual in the $SO(3,C)$ sense.  The Ashtekar formulation arises upon elimination of $\psi_{ae}$ via a new mechanism, which basically restricts one to a functional submanifold of the space of 
actions defined by $I_{Pleb}$.  Using this same mechanism, in section 3 we show that elimination of the two 
forms $\Sigma^a$ in favor of $\psi_{ae}$ yields a new action $I_{Inst}$, the instanton representation of Plebanski gravity.  This shows that $I_{Ash}$ and $I_{Inst}$ are in a sense complementary within $I_{Pleb}$, which suggests that the latter is also a theory of GR.  We prove this rigorously in section 4 by demonstrating that $I_{Inst}$ does indeed reproduce the Einstein equations, combined with a prescription for writing a solution subject to the initial value constraints.\par
\indent 
In section 5 we provide an analysis of the $I_{Inst}$ equations of motion beyond the Einstein equations.  A Hodge duality condition emerges on-shell, which as shown in section 7 explicitly provides the spacetime metric.\footnote{The implication is that the metrics from sections 4 and 7 must be equal to each other as a consistency condition.  This should provide a practical method for constructing GR solutions via what we will name as the instanton representation method.}  In section 6 we clarify the similarities and differences between $I_{Inst}$ and the pure spin connection of CDJ in \cite{SPINCON}.  There are common notions in the community that a certain antecedent of the CDJ action is essentially the same action as $I_{Inst}$.  The present paper shows explicitly that this is not the case and the reasons why, and shows that $I_{Inst}$ is indeed a different action constituting a new approach to GR.  This will as well be independently corroborated by various follow-on papers which apply the instanton representation method.  Section 7 delineates the reality conditions on $I_{Inst}$, which appear to be intimately intertwined with the signature of spacetime.  Sections 8 and 9 clarify a hidden relation of general relativity to Yang--Mills theory, which brings into play the concept of gravitational instantons.\par
\indent
The author has not been able to find, amongst the various sources in the literature, a uniform definition of what a gravitational instanton is.  Some references, for example as in \cite{GRAVINTGAUGE} and \cite{CAPBLANSKI}, define gravitational instantons as GR solutions having a vanishing Weyl tensor with nonvanishing cosmological constant.  This would seem to imply, in the language of the present paper, that gravitational instantons can exist only for spacetimes of Petrov Type O\footnote{The definitions of the various Petrov Types can be found in \cite{PENROSERIND} and in \cite{MACCALLUM}.  The intent of the instanton representation of Plebanski gravity, as presented in this paper, is to be able to classify GR solutions according to their Petrov Type.}  On the other hand, other references (for example \cite{DINSTANT}) allow for Type D gravitational instantons.  In spite of this a common element, barring topological considerations, appears to be that of a solution to the acuum equations having self-dual curvature.  We hope in the present paper to shed some light on the concept of a graviton as a `generalized' Yang--Mills instanton, which can exist as a minimum for Petrov Types I in addition to Types D and O.  Section 10 contains a summary of the main results of this paper and some future directions of research.\par
\indent
On a final note prior to prior to proceeding, we will establish the following index conventions.  Lowercase symbols from the Latin alphabet $a,b,c,\dots$ will denote internal $SO(3,C)$ indices, and those from the middle $i,j,k,\dots$ will denote spatial indices, each taking values $1$, $2$ and $3$.  SL(2,C) indices will be labelled by capital letters $A$ and $A^{\prime}$ taking values $0$ and $1$, and 4-dimensional spacetime 
indices by Greek symbols $\mu,\nu,\dots$.

\section{Plebanski's theory of gravity}

The starting Plebanski action \cite{PLEBANSKI} writes GR using self-dual two forms in lieu of the spacetime metric $g_{\mu\nu}$ as the basic variables.  We adapt the starting action to the language of the $SO(3,C)$ gauge algebra as
\begin{eqnarray}
\label{STARTINGFROM}
I={i \over G}\int_M\delta_{ae}{\Sigma^a}\wedge{F^e}-{1 \over 2}(\delta_{ae}\varphi+\psi_{ae}){\Sigma^a}\wedge{\Sigma^e},
\end{eqnarray}
\noindent
where $\Sigma^a={1 \over 2}\Sigma^a_{\mu\nu}{dx^{\mu}}\wedge{dx^{\nu}}$ are a triplet of $SO(3,C)$ two forms and $F^a={1 \over 2}F^a_{\mu\nu}{dx^{\mu}}\wedge{dx^{\nu}}$ is the field-strength two form 
for gauge connection $A^a=A^a_{\mu}dx^{\mu}$.  Also $\psi_{ae}$ is symmetric and traceless and $\varphi$ is a numerical constant.  The field strength is written in component form 
as $F^a_{\mu\nu}=\partial_{\mu}A^a_{\nu}-\partial_{\nu}A^a_{\mu}+f^{abc}A^b_{\mu}A^c_{\nu}$, with $SO(3,C)$ structure constants $f^{abc}=\epsilon^{abc}$.  The equations of motion resulting from (\ref{STARTINGFROM}) 
are (See e.g. \cite{SELFDUAL} and \cite{KRASNOV})
\begin{eqnarray}
\label{STARTINGFROM3}
{{\delta{I}} \over {\delta{A^g}}}=D\Sigma^g=d\Sigma^g+\epsilon^g_{fh}{A^f}\wedge{\Sigma^h}=0;\nonumber\\
{{\delta{I}} \over {\delta\psi_{ae}}}={\Sigma^a}\wedge{\Sigma^e}-{1 \over 3}\delta^{ae}{\Sigma^g}\wedge{\Sigma_g}=0;\nonumber\\
{{\delta{I}} \over {\delta\Sigma^a}}=F^a-\Psi^{-1}_{ae}\Sigma^e=0\longrightarrow{F}^a_{\mu\nu}=\Psi^{-1}_{ae}\Sigma^e_{\mu\nu}.
\end{eqnarray}
\noindent
The first equation of (\ref{STARTINGFROM3}) states that $A^g$ is the self-dual part of the spin connection compatible with the two forms $\Sigma^a$, where $D$ is the exterior covariant derivative with respect to $A^a$.  The second equation implies that the two forms $\Sigma^a$ can be constructed from tetrad one-forms $e^I=e^I_{\mu}dx^{\mu}$ in the form\footnote{In the tetrad formulation of GR, this corresponds to spacetimes of Lorentzian signature when $e^0$ is real, and Euclidean signature when $e^0$ is pure imaginary.}
\begin{eqnarray}
\label{STARTINGFROM4}
\Sigma^a=i{e^0}\wedge{e^a}-{1 \over 2}\epsilon_{afg}{e^f}\wedge{e^g}.
\end{eqnarray}
\noindent
Equation (\ref{STARTINGFROM4}) is a self-dual combination, which enforces the equivalence of (\ref{STARTINGFROM}) to general relativity.  Note that (\ref{STARTINGFROM4}) implies \cite{KRASNOV}
\begin{eqnarray}
\label{STARTINGFROM41}
{i \over 2}{\Sigma^a}\wedge{\Sigma^e}=\delta^{ae}\sqrt{-g}d^4x,
\end{eqnarray}
\noindent
with the spacetime volume element as the proportionality factor.  The third equation of motion in (\ref{STARTINGFROM3}) states that the curvature of $A^a$ is self-dual as a two form, which implies that the 
metric $g_{\mu\nu}=\eta_{IJ}e^I_{\mu}e^J_{\nu}$ derived from the tetrad one-forms $e^I$ satisfies the vacuum Einstein equations.\par
\indent  
If one were to eliminate the 2-forms $\Sigma^a$ and the matrix $\psi_{ae}$ from the 
action (\ref{STARTINGFROM}) while leaving the connection $A^a_{\mu}$ intact, then one would obtain the CDJ action \cite{SPINCON}, which is written almost completely in terms of the spin connection.  But we would like to obtain a formulation of general relativity which preserves these fields to some extent, since they contain fundamental gravitational degrees of freedom as well as a mechanism for directly implementing the intitial value constraints.\par
\indent  
The most direct way to preserve the ability to implement the constraints in a totally constrained system is to first perform a 3+1 decomposition of the action.  The starting action (\ref{STARTINGFROM}) in component form is given by
\begin{eqnarray}
\label{OPTION1}
I[\Sigma^a,A^a,\Psi]={1 \over 4}\int_Md^4x\Bigl(\Sigma^a_{\mu\nu}F^a_{\rho\sigma}-{1 \over 2}\Psi^{-1}_{ae}\Sigma^a_{\mu\nu}\Sigma^e_{\rho\sigma}\Bigr)\epsilon^{\mu\nu\rho\sigma}
\end{eqnarray}
\noindent
where $\epsilon^{0123}=1$ and we have defined $\Psi^{-1}_{ae}=\delta_{ae}\varphi+\psi_{ae}$.\par
\indent
For $\varphi=-{\Lambda \over 3}$, where $\Lambda$ is the cosmological constant, then we have that
\begin{eqnarray}
\label{OPTION2}
\Psi^{-1}_{ae}=-{\Lambda \over 3}\delta_{ae}+\psi_{ae}.
\end{eqnarray}
\noindent
The matrix $\psi_{ae}$, presented in \cite{CAP} is the self-dual part of the Weyl curvature tensor 
in $SO(3,C)$ language.  The eigenvalues of $\psi_{ae}$ determine the algebraic classification of spacetime which is independent of coordinates and of tetrad frames (\cite{MACCALLUM}, \cite{PENROSERIND}).\footnote{This includes principal null directions and properties of gravitational radiation.}  $\Psi^{-1}_{ae}$ is the matrix inverse of $\Psi_{ae}$ which we will refer to as the CDJ matrix, and is the result of appending to $\psi_{ae}$ a trace part.  In the CDJ formulation this field is eliminated in addition to the 2-forms $\Sigma^a$.\par
\indent
\subsection{The Ashtekar variables}

Assuming a spacetime manifold of topology $M=\Sigma\times{R}$, let us perform a 3+1 decomposition of (\ref{OPTION1}).  
Defining $\widetilde{\sigma}^i_a\equiv{1 \over 2}\epsilon^{ijk}\Sigma^a_{jk}$ and 
${B}^i_a\equiv{1 \over 2}\epsilon^{ijk}F^a_{jk}$ for the spatial parts of the self-dual and curvature two forms, this is given by
\begin{eqnarray}
\label{SECONDTERM7}
I=\int{dt}\int_{\Sigma}d^3x\widetilde{\sigma}^a_i\dot{A}^a_i+A^a_0D_i\widetilde{\sigma}^i_a
+\Sigma^a_{0i}\bigl(B^i_a-\Psi^{-1}_{ae}\widetilde{\sigma}^i_e\bigr),
\end{eqnarray}
\noindent
where we have integrated by parts, using $F^a_{0i}=\dot{A}^a_i-D_iA^a_0$ from the temporal component of the curvature.
We will use (\ref{STARTINGFROM3}) and (\ref{STARTINGFROM4}) to redefine the two form components in (\ref{SECONDTERM7}).  Define $e^a_i$ as the spatial part of the tetrads $e^I_{\mu}$ and make the identification
\begin{eqnarray}
\label{COMPONENT7}
e^a_i={1 \over 2}\epsilon_{ijk}\epsilon^{abc}\widetilde{\sigma}^j_b\widetilde{\sigma}^k_c(\hbox{det}\widetilde{\sigma})^{-1/2}=\sqrt{\hbox{det}\widetilde{\sigma}}(\widetilde{\sigma}^{-1})^a_i.
\end{eqnarray}
\noindent
For a special case $e^0_i=0$, known as the time gauge, then the temporal components of the two forms (\ref{STARTINGFROM4}) are given by (See e.g. \cite{SELFDUAL}, \cite{REISENBERGER})
\begin{eqnarray}
\label{COMPONENT8}
\Sigma^a_{0i}={i \over 2}\underline{N}\epsilon_{ijk}\epsilon^{abc}\widetilde{\sigma}^j_b\widetilde{\sigma}^k_c+\epsilon_{ijk}N^j\widetilde{\sigma}^k_a,
\end{eqnarray}
\noindent
where $\underline{N}=N(\hbox{det}\widetilde{\sigma})^{-1/2}$ with $N$ and $N^i$ being a set of four nondynamical fields.  In the steps leading to the CDJ action, the fields $N^{\mu}=(N,N^i)$ become eliminated along with the process of eliminating the 2-forms $\Sigma^a_{\mu\nu}$.\par
\indent
Substituting (\ref{COMPONENT8}) into (\ref{SECONDTERM7}), we obtain the action
\begin{eqnarray}
\label{SECONDTERM10}
I=\int{dt}\int_{\Sigma}d^3x\widetilde{\sigma}^i_a\dot{A}^a_i+A^a_0G_a-N^iH_i-iNH.
\end{eqnarray}
\noindent
The fields $(A^a_0,N,N^i)$ are auxilliary fields whose variations yield respectively the following constraints
\begin{eqnarray}
\label{GOOSESS}
G_a=D_i\widetilde{\sigma}^i_a;~~H_i=\epsilon_{ijk}\widetilde{\sigma}^j_aB^k_a+\epsilon_{ijk}\widetilde{\sigma}^j_a\widetilde{\sigma}^k_e\Psi^{-1}_{ae};\nonumber\\
H=(\hbox{det}\widetilde{\sigma})^{-1/2}\Bigl({1 \over 2}\epsilon_{ijk}\epsilon^{abc}\widetilde{\sigma}^i_a\widetilde{\sigma}^j_bB^k_c
-{1 \over 6}(\hbox{tr}\Psi^{-1})\epsilon_{ijk}\epsilon_{abc}\widetilde{\sigma}^i_a\widetilde{\sigma}^j_b\widetilde{\sigma}^k_c\Bigr).
\end{eqnarray}
\noindent
Rather than attempt to perform a canonical analysis of (\ref{GOOSESS}), we will proceed from (\ref{SECONDTERM10}) as follows.  Think of $I=I_{\widetilde{\sigma},\Psi}[A]$ as an infinite dimensional functional manifold of theories parametrized by the fields $\widetilde{\sigma}^i_a$ and $\Psi_{ae}$, and then restrict attention to a submanifold corresponding to the theory of GR.\par
\indent 
Following suit, say that we impose the following conditions on $\Psi^{-1}_{ae}$
\begin{eqnarray}
\label{IMPOSE}
\epsilon^{bae}\Psi^{-1}_{ae}=0;~~\hbox{tr}\Psi^{-1}=-\Lambda
\end{eqnarray}
\noindent
with no restrictions on $\widetilde{\sigma}^i_a$, where $\Lambda$ is the cosmological constant.  Then $\Psi^{-1}_{ae}$ becomes eliminated and equation (\ref{SECONDTERM10}) reduces to the action for general relativity in the Ashtekar variables (\cite{ASH1},\cite{ASH2},\cite{ASH3})
\begin{eqnarray}
\label{ASHTESH}
I_{Ash}={1 \over G}\int{dt}\int_{\Sigma}d^3x\widetilde{\sigma}^i_a\dot{A}^a_i+A^a_0D_i\widetilde{\sigma}^i_a\nonumber\\
-\epsilon_{ijk}N^i\widetilde{\sigma}^j_aB^k_a+{i \over 2}\underline{N}\epsilon_{ijk}\epsilon_{abc}\widetilde{\sigma}^i_a\widetilde{\sigma}^j_b\bigl(B^k_c+{\Lambda \over 3}\widetilde{\sigma}^k_c\bigr),
\end{eqnarray}
\noindent
where $\underline{N}=N(\hbox{det}\widetilde{\sigma})^{-1/2}$ is the lapse density function.  The action (\ref{ASHTESH}) is written on the phase space $\Omega_{Ash}=(\widetilde{\sigma}^i_a,A^a_i)$ and the variable $\Psi^{-1}_{ae}$ has been eliminated.  The auxilliary fields $A^a_0$, $N$ and $N^i$ respectively 
are the $SO(3,C)$ rotation angle, the lapse function and the shift vector.  The auxilliary fields are Lagrange multipliers smearing their corresponding initial value constraints $G_a$, $H$ and $H_i$, respectively the Gauss' law, Hamiltonian and diffeomorphism constraints.  Note that $\widetilde{\sigma}^i_a$ in the original Plebanski action was part of an auxilliary field $\Sigma^a_{\mu\nu}$, but now in (\ref{ASHTESH}) it has been promoted to the status of a momentum space dynamical variable.  At the level of (\ref{ASHTESH}), one could further eliminate the 2-forms $\Sigma^a$ to obtain the CDJ pure spin connection action \cite{SPINCON}.  However, (\ref{ASHTESH}) is already in a form suitable to quantization and implementation of the initial value constraints via the temporal parts of these 2-forms.

\section{The instanton representation}

Having shown that Plebanski's action (\ref{STARTINGFROM}) contains (\ref{ASHTESH}), an action known to describe GR, as a direct consequence of (\ref{IMPOSE}), we will now show that (\ref{STARTINGFROM}) also contains an alternate formulation of GR based on the field $\Psi_{ae}$, which can also be derived directly from (\ref{OPTION1}).  Instead of equation (\ref{IMPOSE}), let us impose the following conditions in (\ref{GOOSESS})
\begin{eqnarray}
\label{OBTAINAN1}
\epsilon_{ijk}\epsilon_{abc}\widetilde{\sigma}^i_a\widetilde{\sigma}^j_bB^k_c=-{\Lambda \over 3}\epsilon_{ijk}\epsilon_{abc}\widetilde{\sigma}^i_a\widetilde{\sigma}^j_b\widetilde{\sigma}^k_c;~~
\epsilon_{ijk}\widetilde{\sigma}^j_aB^k_a=0
\end{eqnarray}
\noindent
with no restriction on $\Psi_{ae}$.  Substitution of (\ref{OBTAINAN1}) into (\ref{GOOSESS}) yields
\begin{eqnarray}
\label{OBTAINAN2}
H_i=\epsilon_{ijk}\widetilde{\sigma}^j_a\widetilde{\sigma}^k_e\Psi^{-1}_{ae};\nonumber\\
H=(\hbox{det}\widetilde{\sigma})^{-1/2}\Bigl(-{\Lambda \over 6}\epsilon_{ijk}\epsilon_{abc}\widetilde{\sigma}^i_a\widetilde{\sigma}^j_b\widetilde{\sigma}^k_c\nonumber\\
-{1 \over 6}(\hbox{tr}\Psi^{-1})\epsilon_{ijk}\epsilon_{abc}\widetilde{\sigma}^i_a\widetilde{\sigma}^j_b\widetilde{\sigma}^k_c\Bigr)
=-\sqrt{\hbox{det}\widetilde{\sigma}}\bigl(\Lambda+\hbox{tr}\Psi^{-1}\bigr).
\end{eqnarray}
\noindent
Hence substituting (\ref{OBTAINAN2}) into (\ref{SECONDTERM10}), we obtain an action given by
\begin{eqnarray}
\label{ISGIVEN}
I=\int{dt}\int_{\Sigma}d^3x\widetilde{\sigma}^i_a\dot{A}^a_i+A^a_0D_i\widetilde{\sigma}^i_a\nonumber\\
+\epsilon_{ijk}N^i\widetilde{\sigma}^j_a\widetilde{\sigma}^k_e\Psi^{-1}_{ae}-iN\sqrt{\hbox{det}\widetilde{\sigma}}\bigl(\Lambda+\hbox{tr}\Psi^{-1}\bigr).
\end{eqnarray}
\noindent
But (\ref{ISGIVEN}) still contains $\widetilde{\sigma}^i_a$, therefore we will completely eliminate $\widetilde{\sigma}^i_a$ by substituting the spatial restriction of the third equation of motion of (\ref{STARTINGFROM3})
\begin{eqnarray}
\label{ISGIVEN1}
\widetilde{\sigma}^i_a=\Psi_{ae}B^i_e,
\end{eqnarray}
\noindent
into (\ref{ISGIVEN}).  This substitution, which also appears in \cite{SPINCON} in the form of the so-called CDJ Ansatz, yields the action\footnote{Equation (\ref{ISGIVEN1}) is valid $B^i_a$ and $\Psi_{ae}$ are nondegenerate as three by three matrices.  Hence all results of this paper will be confined to configurations where this is the case.}
\begin{eqnarray}
\label{ISGIVEN2}
I_{Inst}=\int{dt}\int_{\Sigma}d^3x\Psi_{ae}B^i_a\dot{A}^a_i+A^a_0B^i_eD_i\Psi_{ae}\nonumber\\
+\epsilon_{ijk}N^iB^j_aB^k_e\Psi_{ae}-iN(\hbox{det}B)^{1/2}\sqrt{\hbox{det}\Psi}\bigl(\Lambda+\hbox{tr}\Psi^{-1}\bigr),
\end{eqnarray}
\noindent
which depends on the CDJ matrix $\Psi_{ae}$ and the Ashtekar connection $A^a_i$, with no appearance of $\widetilde{\sigma}^i_a$.  In the original Plebanski theory $\Psi_{ae}$ was an auxilliary field which could be eliminated.  But now $\Psi_{ae}$ has been promoted to the status of a full dynamical variable, analogously to $\widetilde{\sigma}^i_a$ in $I_{Ash}$.\par
\indent
There are a few items of note regarding (\ref{ISGIVEN2}).  Note that it contains the same auxilliary fields $(A^a_0,N,N^i)$ as in the Ashtekar theory.  Since we have imposed the constraints $H_{\mu}=(H,H_i)$ on the Ashtekar phase space within the starting Plebanski theory in order to obtain $I_{Inst}$, then this implies that the initial value constraints $(G_a,H,H_i)$ must play an anlogous role in (\ref{ISGIVEN2}) as their counterparts in (\ref{ASHTESH}).  This relation holds only where $\Psi_{ae}$ is nondegenerate, which limits one to spacetimes of Petrov Type I, D and O where $\Psi_{ae}$ has three linearly independent eigenvectors.\footnote{We refer to (\ref{ISGIVEN2}) as the instanton representation of Plebanski gravity because it follows directly from Plebanski's action (\ref{STARTINGFROM}).  We will use in this sense use (\ref{ISGIVEN2}) as the starting point for the reformulation of gravity as presented.  The association of (\ref{ISGIVEN2}) to gravitational instantons will be made more precise later in this paper.}  Lastly, note that by further elimination of $\Psi_{ae}$ and $N^i$ from (\ref{ISGIVEN2}) one can obtain the CDJ action in \cite{SPINCON}.  However, we would like to preserve $\Psi_{ae}$ since it contains gravitational degrees of freedom relevant to the instanton representation, and the shift vector $N^i$ as we will see also assumes an important role. 

\section{Equations of motion from the instanton representation}

We will now show that Einstein equations follow from the instanton representation action (\ref{ISGIVEN2}) in the same sense that they follow from the original Plebanski action (\ref{STARTINGFROM}).  More precisely, we will demonstrate consistency of the equations of motion with equations (\ref{STARTINGFROM3}) and (\ref{STARTINGFROM4}).  After integrating by parts and discarding boundary terms, the starting action (\ref{ISGIVEN2}) is given by 
\begin{eqnarray}
\label{OPTION78}
I_{Inst}=\int{dt}\int_{\Sigma}d^3x\Psi_{ae}B^k_e\bigl(F^a_{0i}+\epsilon_{kjm}B^j_aN^m\bigr)\nonumber\\
-iN(\hbox{det}B)^{1/2}\sqrt{\hbox{det}\Psi}\bigl(\Lambda+\hbox{tr}\Psi^{-1}\bigr).
\end{eqnarray}
\noindent
The equation of motion for the shift vector $N^i$ is given by
\begin{eqnarray}
\label{SHIFT}
{{\delta{I}_{Inst}} \over {\delta{N}^i}}=\epsilon_{mjk}B^j_aB^k_e\Psi_{ae}=0,
\end{eqnarray}
\noindent
which implies on the solution to the equations of motion that $\Psi_{ae}=\Psi_{(ae)}$ is symmetric.  The equation of motion for the lapse function $N$ is given by
\begin{eqnarray}
\label{OPTION76}
{{\delta{I}_{Inst}} \over {\delta{N}}}=(\hbox{det}B)^{1/2}\sqrt{\hbox{det}\Psi}\bigl(\Lambda+\hbox{tr}\Psi^{-1}\bigr)=0.
\end{eqnarray}
\noindent
Nondegeneracy of $\Psi_{ae}$ and of the magnetic field $B^i_e$ implies that on-shell, the following relation must be satisfied
\begin{eqnarray}
\label{HAMM}
\Lambda+\hbox{tr}\Psi^{-1}=0,
\end{eqnarray}
\noindent
which implies that $\lambda_3$ can be written explicitly in terms of $\lambda_1$ and $\lambda_2$, regarded as physical degrees of freedom.  The equation of motion for $\Psi_{ae}$ is 
\begin{eqnarray}
\label{HODGE}
{{\delta{I}_{Inst}} \over {\delta\Psi_{ae}}}=B^k_eF^a_{0k}+\epsilon_{kjm}B^k_eB^j_aN^m+iN\sqrt{\hbox{det}B}\sqrt{\hbox{det}\Psi}(\Psi^{-1}\Psi^{-1})^{ea}=0,
\end{eqnarray}
\noindent
where we have used (\ref{HAMM}).  The symmetric and the antisymmetric parts of (\ref{HODGE}) must separately vanish.  The antisymmetric part is given by 
\begin{eqnarray}
\label{HODGE1}
B^k_{[e}F^{a]}_{0k}+\epsilon_{mkj}N^mB^k_eB^j_a=0,
\end{eqnarray}
\noindent
which can be used to solve for the shift vector $N^i$.  Using the relation $\epsilon_{ijk}B^j_aB^k_e=\epsilon_{aed}(B^{-1})^d_i(\hbox{det}B)$ for nondegenerate 3 by 3 matrices, we have
\begin{eqnarray}
\label{HODGE2}
N^i={1 \over 2}\epsilon^{ijk}F^g_{0j}(B^{-1})^g_k.
\end{eqnarray}
\noindent
The symmetric part of (\ref{HODGE}) is given by
\begin{eqnarray}
\label{HODGE3}
B^k_{(e}F^{a)}_{0k}+iN\sqrt{\hbox{det}B}\sqrt{\hbox{det}\Psi}(\Psi^{-1}\Psi^{-1})^{(ea)}=0,
\end{eqnarray}
\noindent
where we have used that $\Psi_{ae}$ on-shell is symmetric from (\ref{SHIFT}).\par
\indent  

\subsection{Proof of the Einstein equations}

To make a direct connection from the instanton representation to Einstein's general relativity, we will show that the equations of motion for $I_{Inst}$ imply the Einstein equations.  Let us use the relation
\begin{eqnarray}
\label{USING}
\sqrt{-g}=N\sqrt{h}=N\sqrt{\hbox{det}\widetilde{\sigma}}=(\hbox{det}B)^{1/2}\sqrt{\hbox{det}\Psi},
\end{eqnarray}
\noindent
which writes the determinant of the spacetime metric $g_{\mu\nu}$ in terms of dynamical variables $(A,\Psi)$ using the 3+1 decomposition and uses the determinant of (\ref{ISGIVEN1}).  Defining $\epsilon^{0ijk}\equiv\epsilon^{ijk}$ and using the symmetries of the four dimensional epsilon tensor $\epsilon^{\mu\nu\rho\sigma}$, then the following identities hold
\begin{eqnarray}
\label{HODGE4}
B^k_{(e}F^{a)}_{0k}={1 \over 2}\epsilon^{klm}F^{(e}_{lm}F^{a)}_{0k}={1 \over 8}\epsilon^{\mu\nu\rho\sigma}F^a_{\mu\nu}F^e_{\rho\sigma}.
\end{eqnarray}
\noindent
Using (\ref{HODGE4}) and (\ref{USING}), then equation (\ref{HODGE3}) can be re-written as
\begin{eqnarray}
\label{OPTION761}
{1 \over 8}F^b_{\mu\nu}F^f_{\rho\sigma}\epsilon^{\mu\nu\rho\sigma}+i\sqrt{-g}(\Psi^{-1}\Psi^{-1})^{(bf)}=0.
\end{eqnarray}
\noindent
Left and right multiplying (\ref{OPTION761}) by $\Psi$, which is symmetric after implementation of (\ref{SHIFT}), we obtain
\begin{eqnarray}
\label{OPTION762}
{1 \over 4}(\Psi^{bb^{\prime}}F^{b^{\prime}}_{\mu\nu})(\Psi^{ff^{\prime}}F^{f^{\prime}}_{\rho\sigma})\epsilon^{\mu\nu\rho\sigma}=-2i\sqrt{-g}\delta^{bf}.
\end{eqnarray}
\noindent
Note that this step and the steps that follow require that $\Psi_{ae}$ be nondegenerate as a 3 by 3 matrix.  Let us make the definition
\begin{eqnarray}
\label{RECOGNIZES}
\Sigma^a_{\mu\nu}=\Psi_{ae}F^e_{\mu\nu}=\Sigma^a_{\mu\nu}[\Psi,A],
\end{eqnarray}
\noindent
which retains $\Psi_{ae}$ and $A^a_{\mu}$ as fundamental, with the two form $\Sigma^a_{\mu\nu}$ being derived quantities.  Upon using the third line of (\ref{STARTINGFROM3}) as a re-definition of variables, which amounts to using the curvature and the CDJ matrix to construct a two form, (\ref{OPTION762}) reduces to
\begin{eqnarray}
\label{REDEF}
{1 \over 4}\Sigma^b_{\mu\nu}\Sigma^f_{\rho\sigma}\epsilon^{\mu\nu\rho\sigma}{dx^{\mu}}\wedge{dx^{\nu}}\wedge{dx^{\rho}}\wedge{dx^{\sigma}}={\Sigma^b}\wedge{\Sigma^f}=-2i\sqrt{-g}\delta^{bf}d^4x.
\end{eqnarray}
\noindent
One recognizes (\ref{REDEF}) as the condition that the two forms thus constructed, which are now derived quantities, be derivable from tetrads, which is the analogue of  (\ref{STARTINGFROM41}).  Indeed, one can conclude as a consequence of (\ref{REDEF}) that there exist one forms $\theta^I=\theta^I_{\mu}dx^{\mu}$ where $I=0,1,\cdots{4}$, such that
\begin{eqnarray}
\label{RECOGNIZES}
\Psi_{ae}F^e=i{\theta^0}\wedge{\theta^a}-{1 \over 2}\epsilon_{afg}{\theta^f}\wedge{\theta^g}\equiv{P}^a_{fg}{\theta^f}\wedge{\theta^g}.
\end{eqnarray}
\noindent
We have defined $P^a_{fg}$ as a projection operator onto the self-dual combination of one-form wedge products, self-dual in the $SO(3,C)$ sense.  To complete the demonstration that the instanton representation yields the Einstein equations, it remains to show that the connection $A^a$ is compatible with the two forms $\Sigma^a$ as constructed in (\ref{RECOGNIZES}).\par
\indent
Using the fact that $\Psi_{ae}$ is symmetric on solutions to (\ref{SHIFT}), the starting action (\ref{OPTION78}) can be written as\footnote{The same action was written down in \cite{SPINCON}, which arises from eliminating the self-dual 2-forms directly from Plebanski's action.  In the approach of the present paper, we have eliminated only the spatial part of the 2-forms, and have used the antisymmetric part of $\Psi_{ae}$ to solve for the shift vector.}
\begin{eqnarray}
\label{OPTION78N}
I_{Inst}=\int_Md^4x{1 \over 8}\Psi_{ae}F^a_{\mu\nu}F^e_{\rho\sigma}\epsilon^{\mu\nu\rho\sigma}-i\sqrt{-g}\bigl(\Lambda+\hbox{tr}\Psi^{-1}\bigr).
\end{eqnarray}
\noindent
The equation of motion for the connection $A^a_{\mu}$ from (\ref{OPTION78N}) is given by
\begin{eqnarray}
\label{MOTION11}
{{\delta{I}_{Inst}} \over {\delta{A}^a_{\mu}}}
\sim\epsilon^{\mu\sigma\nu\rho}D_{\sigma}(\Psi_{ae}F^e_{\nu\rho})
-{i \over 2}\delta^{\mu}_iD^{ij}_{da}\Bigl(N(B^{-1})^d_j\sqrt{\hbox{det}B}\sqrt{\hbox{det}\Psi}\bigl(\Lambda+\hbox{tr}\Psi^{-1}\bigr)\Bigr),
\end{eqnarray}
\noindent
where we have used that $\Psi_{ae}$ is symmetric and we have defined
\begin{eqnarray}
\label{DECFIN}
\overline{D}^{ji}_{ea}(x,y)\equiv{\delta \over {\delta{A}^a_i(x)}}B^j_e(y)=\epsilon^{jki}\bigl(-\delta_{ae}\partial_k+f_{eda}A^d_k\bigr)\delta^{(3)}(x,y);~~\overline{D}^{0i}_{ea}\equiv{0}.
\end{eqnarray}
\noindent
The term in brackets in (\ref{MOTION11}) vanishes since it is proportional to the equation of motion (\ref{OPTION76}) and its spatial derivatives, which leaves us with
\begin{eqnarray}
\label{MOTION111}
\epsilon^{\mu\sigma\nu\rho}D_{\sigma}(\Psi_{ae}F^e_{\nu\rho})=0.
\end{eqnarray}
\noindent
Equation (\ref{MOTION111}) states that when (\ref{SHIFT}) and (\ref{HAMM}) are satisfied, then the two forms $\Sigma^a_{\mu\nu}$ constructed from $\Psi_{ae}$ and $F^e_{\mu\nu}$ as in (\ref{RECOGNIZES}) are compatible with the connection $A^a_{\mu}$.  This is the direct analogue of the first equation from (\ref{STARTINGFROM3}).\par
\indent
Using (\ref{OPTION78}) as the starting point, which uses $\Psi_{ae}$ and $A^a_{\mu}$ as the dynamical variables, we have obtained the Einstein equations in the same sense as the starting Plebanski theory (\ref{STARTINGFROM}) implies the Einstein equations.  The first equation of (\ref{STARTINGFROM3}) has been reproduced via (\ref{MOTION111}), which holds provided that (\ref{HAMM}) and (\ref{SHIFT}) are satisfied.  The second equation of (\ref{STARTINGFROM3}) has been reproduced via (\ref{REDEF}), which follows from (\ref{OPTION761}) when (\ref{SHIFT}) is satisfied.  The third equation of (\ref{STARTINGFROM3}) may be regarded as a defining relation for the instanton representation.  Since the Einstein equations have arisen from the instanton representation, then it follows that $I_{Inst}$ is another representation for general relativity for nondegenerate $\Psi_{ae}$ and $B^i_e$.\par
\indent
On the solution to (\ref{SHIFT}) and (\ref{HAMM}) and using (\ref{RECOGNIZES}), the action for the instanton representation can be written in the language of two forms as 
\begin{eqnarray}
\label{ELIMINATE4}
I_{Inst}={1 \over 2}\int_M\Psi_{bf}{F^b}\wedge{F^f}={1 \over 2}\int_MP^a_{fg}{\theta^f}\wedge{\theta^g}\wedge{F^e},
\end{eqnarray}
\noindent
which upon the identification of one forms $\theta^I$ with tetrads, is nothing other than the self-dual Palatini action \cite{PELDAN}.  Note that the Palatini action implies the Einstein equations with respect to the metric defined by
\begin{eqnarray}
\label{ELIMINATE5}
ds^2=g_{\mu\nu}{dx^{\mu}}{dx^{\nu}}=\eta_{IJ}\theta^I\otimes\theta^J,
\end{eqnarray}
\noindent
where $\eta_{IJ}$ is the Minkowski metric, which provides additional confirmation that the instanton representation $I_{Inst}$ describes Einstein's general relativity when $\Psi_{ae}$ is nondegenerate.

\subsection{Discussion: Constructing a solution}

We have shown how the Einstein equations follow from the instanton representation (\ref{ISGIVEN2}), which uses $\Psi_{ae}$ and $A^a_{\mu}$ as the dynamical variables.  Equation (\ref{OPTION762}) implies the existence of a tetrad, which imposes the equivalence to general relativity, but it does not explain how to construct the tetrad.  Since the spacetime metric $g_{\mu\nu}$ is the fundamental variable in Einstein's theory, we will bypass the tetrad and construct $g_{\mu\nu}$ directly as follows.  Perform a 3+1 decomposition of spacetime $M=\Sigma\times{R}$, where $\Sigma$ is a 3-dimensional spatial hypersurface.  The line element is given by 
\begin{eqnarray}
\label{DECOMP}
ds^2=g_{\mu\nu}dx^{\mu}dx^{\nu}
=-N^2dt^2+h_{ij}\omega^i\otimes\omega^j,
\end{eqnarray}
\noindent
where $h_{ij}$ is the induced 3-metric on $\Sigma$, and we have defined the one form
\begin{eqnarray}
\label{DECOMP1}
\omega^i=dx^i-N^idt.
\end{eqnarray}
\noindent
The shift vector is given by (\ref{HODGE2}), rewritten here for completeness
\begin{eqnarray}
\label{HODGE21}
N^i={1 \over 2}\epsilon^{ijk}F^g_{0j}(B^{-1})^g_k,
\end{eqnarray}
and the lapse function $N$ can apparently be chosen freely.  To complete the construction of $g_{\mu\nu}$ using (\ref{OPTION76}) as the starting point we must write the 3-metric $h_{ij}$ using $\Psi_{ae}$ and $A^a_{\mu}$.  The desired expression is given by
\begin{eqnarray}
\label{METRIC}
h_{ij}=(\hbox{det}\Psi)(\Psi^{-1}\Psi^{-1})^{ae}(B^{-1})^a_i(B^{-1})^e_j(\hbox{det}B)=h_{ij}[\Psi,A],
\end{eqnarray}
\noindent
where the following conditions must be satisfied
\begin{eqnarray}
\label{DECOMP3}
\textbf{w}_e\{\Psi_{ae}\}=B^i_eD_i\Psi_{ae}=0;~~\epsilon_{dae}\Psi_{ae}=0;~~\Lambda+\hbox{tr}\Psi^{-1}=0.
\end{eqnarray}
\noindent
Equations (\ref{DECOMP3}) will be referred to as the Gauss' law, diffeomorphism and Hamiltonian constraints, which follow from variation of $A^a_0$, $N^i$ and $N$ in (\ref{ISGIVEN2}).  Note that equation (\ref{DECOMP3}) invovle only  $\Psi_{ae}$ and the spatial spart of the $A^a_{\mu}$, objects which determine a spatial metric in (\ref{METRIC}).  The spacetime metric $g_{\mu\nu}$ solving the Einstein equations is given by
\begin{displaymath}
g_{\mu\nu}=
\left(\begin{array}{cc}
-N^2+N^iN_i & -N_i\\
-N_i & h_{ij}\\
\end{array}\right)
.
\end{displaymath}
\noindent
There are a few things to note regarding this.  (i) From (\ref{HODGE21}), the shift vector $N^i$ depends only on $A^a_{\mu}$, which contains gauge degrees of freedom in the temporal component $A^a_0$. (ii) Secondly, the lapse function $N$ is freely specifiable. (iii) Third, each $A^a_i$ and $\Psi_{ae}$ satisfying the initial value constraints (\ref{DECOMP3}) determines 
a 3-metric $h_{ij}$, which when combined with a choice of $A^a_0$ and $N$ should provide a solution for spacetimes of Petrov Type I, D and O.\par
\indent
Note, when one uses the CDJ Ansatz $\widetilde{\sigma}^i_a=\Psi_{ae}B^i_e$ that (\ref{METRIC}) implies $hh^{ij}=\widetilde{\sigma}^i_a\widetilde{\sigma}^j_a$, which is the relation of the Ashtekar densitized triad to the contravariant 3-metric $h^{ij}$.\footnote{It is shown explicitly in \cite{EYO} that this defines the spatial part of a spacetime metric $g_{\mu\nu}$ solving the Einstein equations.}  Upon implementation 
of (\ref{DECOMP3}) on the phase space $\Omega_{Inst}$, then one is left with the two degrees of freedom per point of GR, and $h_{ij}$ is expressed explicitly in terms of these degrees of freedom.\par
\indent

\section{Analysis of the equations of motion}

We will now provide a rudimentary analysis of the physical content of the equations of motion of $I_{Inst}$ beyond the Einstein equations.  The first equation of motion of $I_{Inst}$ is (\ref{HODGE}) re-written here for completeness
\begin{eqnarray}
\label{DECOMPOSE}
B^i_fF^b_{0i}+i\sqrt{-g}(\Psi^{-1}\Psi^{-1})^{fb}+\epsilon_{ijk}B^i_fB^j_bN^k=0.
\end{eqnarray}
\noindent
Also, when (\ref{SHIFT}) and (\ref{HAMM}) are satisfied, then (\ref{MOTION11}) implies (\ref{MOTION111}), also written here
\begin{eqnarray}
\label{SHOODGEE1}
\epsilon^{\mu\sigma\nu\rho}D_{\sigma}(\Psi_{ae}F^e_{\nu\rho})=0.
\end{eqnarray}
\noindent
We have shown that when $\Psi_{bf}$ is symmetric after determination of $N^i$ as in (\ref{HODGE2}), that the symmetric part of (\ref{DECOMPOSE}) in conjunction with (\ref{SHOODGEE1}) imply the Einstein equations.  We will now show under this condition that (\ref{DECOMPOSE}) and (\ref{SHOODGEE1}) form a self-consistent system.  Act on (\ref{SHOODGEE1}) with $D_{\mu}$ and use the definition of curvature as the commutator of covariants derivatives, yielding
\begin{eqnarray}
\label{SHOODGEE2}
\epsilon^{\mu\nu\rho\sigma}D_{\mu}D_{\nu}(\Psi_{ae}F^e_{\rho\sigma})=f_{abc}\Psi_{ce}(F^b_{\mu\nu}F^e_{\rho\sigma}\epsilon^{\mu\nu\rho\sigma})=0.
\end{eqnarray}
\noindent
Then substituting the symmetric part of (\ref{DECOMPOSE}) into (\ref{SHOODGEE2}), up to an insignificant numerical factor we get
\begin{eqnarray}
\label{SHOODGEE3}
f_{abc}\Psi_{ce}(i\sqrt{-g}(\Psi^{-1}\Psi^{-1})^{(eb)})=i\sqrt{-g}f_{abc}(\Psi^{-1})^{cb}=0,
\end{eqnarray}
\noindent
which is simply the statement that $\Psi_{ce}$ is symmetric in $c$ and $e$ which is consistent with (\ref{SHIFT}) for $(\hbox{det}B)\neq{0}$.  This can also be seen at the level of 2-forms by elimination of 
the curvature from (\ref{SHOODGEE2}) to obtain
\begin{eqnarray}
\label{CANBESEEN}
f_{abf}F^b_{\mu\nu}\Sigma^c_{\rho\sigma}\epsilon^{\mu\nu\rho\sigma}\longrightarrow{f}_{abc}(\Psi^{-1})^{bf}\Sigma^f_{\mu\nu}\Sigma^c_{\rho\sigma}\epsilon^{\mu\nu\rho\sigma}\sim{0}
\end{eqnarray}
\noindent
due to (\ref{RECOGNIZES}), on account of antisymmetry of the structure constants. 
\par
\indent
We will now multiply (\ref{DECOMPOSE}) by $(B^{-1})^f_k$, in conjunction with using the identity $(B^{-1})^d_jB^j_b=\delta^d_b$ since $B^i_f$ is nondegenerate.  Then equation (\ref{DECOMPOSE}) can be written as
\begin{eqnarray}
\label{DECOMPOSE5}
F^b_{0k}+iN(\hbox{det}B)^{-1/2}(\hbox{det}\Psi)^{-1/2}[(\hbox{det}B)(\hbox{det}\Psi)]\bigl((\Psi^{-1}\Psi^{-1})^{df}(B^{-1})^d_j(B^{-1})^f_k\bigr)B^j_b\nonumber\\
+\epsilon_{kjm}B^j_bN^m=0.
\end{eqnarray}
\noindent
We can now use (\ref{METRIC}) in the second term of (\ref{DECOMPOSE5}), which defines the spatial 3-metric in terms of $\Psi_{ae}$ and the spatial connection $A^a_i$ solving the constraints (\ref{DECOMP3}).  Using this in conjunction with the 
relation $N(\hbox{det}B)^{-1/2}(\hbox{det}\Psi)^{-1/2}=Nh^{-1/2}=\underline{N}$, then equation (\ref{DECOMPOSE5}) becomes
\begin{eqnarray}
\label{DECOMPOSE7}
F^b_{0i}+i\underline{N}h_{ij}B^j_b+\epsilon_{ijk}B^j_bN^k=0.
\end{eqnarray}
\noindent
We will show in the next subsection that (\ref{DECOMPOSE7}) is simply the statement that $F^a_{\mu\nu}$ is Hodge self-dual with respect to a metric $g_{\mu\nu}$ whose spatial part is $h_{ij}$, whose lapse function is $N$ and whose shift vector is $N^i$.\par
\indent  
It may appear via (\ref{OPTION762}) that only the symmetric part of (\ref{DECOMPOSE}) is needed in order for $I_{Inst}$ to imply the Einstein equations for Petrov Types I, D and O.  But we have utilized the equation 
of motion (\ref{DECOMPOSE}) to arrive at (\ref{DECOMPOSE7}), which includes information derived using the antisymmetric part of $\Psi_{ae}$.  The reconciliation is in the observation that part of the process of solving the Einstein equations involves computing the shift vector via (\ref{HODGE2}), which simultaneously eliminates the antisymmetric part of (\ref{DECOMPOSE}).  Since (\ref{DECOMPOSE7}) then is consistent with the Einstein equations, then the implication is that each such solution is included within the class of configurations under which the curvature $F^a_{\mu\nu}$ is Hodge self-dual with respect to the corresponding metric $g_{\mu\nu}$.  The spatial part $h_{ij}$ of this metric is defined on the configurations $(\Psi_{ae},A^a_i)$ satisfying (\ref{DECOMP3}).
\noindent

\subsection{Dynamical Hodge self-duality operator}

We will now prove that equation (\ref{DECOMPOSE7}) is indeed the statement that the curvature $F^a_{\mu\nu}$ is Hodge self-dual with respect to $g_{\mu\nu}=g_{\mu\nu}[\Psi,A]$.  To show this, we will derive the the Hodge self-duality condition for Yang--Mills theory in curved spacetime, using the 3+1 decomposition of the associated metric.  The following relations will be useful 
\begin{eqnarray}
\label{SELFISH}
g^{00}=-{1 \over {N^2}};~~g^{0i}=-{{N^i} \over {N^2}};~~g^{ij}=h^{ij}-{{N^iN^j} \over {N^2}},
\end{eqnarray}
where $N$ is real for Lorentzian signature spacetimes and pure imaginary for Euclidean signature.  The Hodge self-duality condition for $F^a_{\mu\nu}$ can be written in the form
\begin{eqnarray}
\label{SELFISH1}
\sqrt{-g}g^{\mu\rho}g^{\nu\sigma}F^a_{\rho\sigma}={\beta \over 2}\epsilon^{\mu\nu\rho\sigma}F^a_{\rho\sigma},
\end{eqnarray}
\noindent
where $\beta$ is a numerical constant which we will determine.  Expanding (\ref{SELFISH1}) and using $F^a_{00}=0$, we have
\begin{eqnarray}
\label{SELFISH2}
N\sqrt{h}\bigl(\bigl(g^{\mu0}g^{\nu{j}}-g^{\nu0}g^{\mu{j}}\bigr)F^a_{0j}+g^{\mu{i}}g^{\nu{j}}\epsilon_{ijk}B^k_a\bigr)
={\beta \over 2}\bigl(2\epsilon^{\mu\nu{0}i}F^a_{0i}+\epsilon^{\mu\nu{ij}}\epsilon_{ijm}B^m_a\bigr).
\end{eqnarray}
\noindent
We will now examine the components of (\ref{SELFISH2}).  The $\mu=0,\nu=0$ component yields $0=0$, which is trivially satisfied.  Moving on to the $\mu=0,\nu=k$ component, we have
\begin{eqnarray}
\label{SELFISH3}
N\sqrt{h}\bigl(\bigl(g^{00}g^{kj}-g^{k0}g^{0j}\bigr)F^a_{0j}+g^{0i}g^{kj}\epsilon_{ijm}B^m_a\bigr)=\beta{B}^k_a.
\end{eqnarray}
\noindent
Making use of (\ref{SELFISH}) as well as the antisymmetry of the epsilon symbol, after some algebra (see Appendix A leading to equation (\ref{FROMTHE3})) we obtain
\begin{eqnarray}
\label{SELFISH4}
F^a_{0j}+\epsilon_{jmk}B^m_aN^k+\beta\underline{N}h_{jk}B^k_a=0,
\end{eqnarray}
\noindent
where we have defined $\underline{N}=N/\sqrt{h}$.  Note that (\ref{SELFISH4}) is the same as (\ref{DECOMPOSE7}) for $\beta=i$, which establishes Hodge self-duality with respect to the spatio-temporal components.  We must next verify  Hodge duality with respect to the purely spatial components of the curvature.  For the $\mu=m,\nu=n$ component of (\ref{SELFISH1}), we have
\begin{eqnarray}
\label{SELFISH5}
N\sqrt{h}\bigl(\bigl(g^{m0}g^{nj}-g^{n0}g^{mj}\bigr)F^a_{0j}+g^{mi}g^{nj}\epsilon_{ijk}B^k_a\bigr)=\beta\epsilon^{mn0i}F^a_{0i}.
\end{eqnarray}
\noindent
Substitution of (\ref{SELFISH}) into (\ref{SELFISH5}) after some algebra (see Appendix A leading to equation (\ref{FROMTHE11})) yields
\begin{eqnarray}
\label{SELFISH6}
{{\sqrt{h}} \over N}\bigl(N^nh^{mj}-N^mh^{nj}\bigr)\bigl(F^a_{0j}+\epsilon_{jkl}B^k_aN^l\bigr)=\epsilon^{mnl}\bigl(\beta{F}^a_{0l}-\underline{N}h_{lk}B^k_a\bigr).
\end{eqnarray}
\noindent
Using $h^{ij}h_{jk}=\delta^i_k$ and simplifying, then (\ref{SELFISH6}) reduces to
\begin{eqnarray}
\label{SELFISH7}
F^a_{0k}+\epsilon_{kmn}B^m_aN^n={1 \over \beta}\underline{N}h_{kl}B^l_a.
\end{eqnarray}
\noindent
Consistency of (\ref{SELFISH7}) with (\ref{SELFISH4}) implies that ${1 \over \beta}=-\beta$, or $\beta=\pm{i}$.\par
\indent
Comparison of (\ref{SELFISH4}) and (\ref{SELFISH7}) with (\ref{DECOMPOSE7}) shows that the Hodge self-duality condition arises dynamically from the equations of motion (\ref{ISGIVEN2}) of $I_{Inst}$.  Moreover, the 
curvature $F^a_{\mu\nu}$ is Hodge self-dual with respect to this operator, which can be written as\footnote{It appears that $\beta=\pm{i}$ follows from our choice of a Lorentzian signature metric corresponding to real $N$, and that one can make a Wick rotation $N\rightarrow{i}N$, and analogously require $\beta=\pm{1}$ for Euclidean signature.  However, we will show in this paper that the signature of spacetime if determined by reality conditions and not by the choice of lapse function $N$.}
\begin{eqnarray}
\label{HODGEIT}
H_{\pm}^{\mu\nu\rho\sigma}={1 \over 2}\bigl(\sqrt{-g}\bigl(g^{\mu\rho}g^{\nu\sigma}-g^{\nu\rho}g^{\mu\sigma}\bigr)\pm{i}\epsilon^{\mu\nu\rho\sigma}\bigr),
\end{eqnarray}
\noindent
where $g_{\mu\nu}=g_{\mu\nu}[\Psi,A]$ is defined in terms of instanton representation variables.  The results can then be summarized as follows.  The instanton representation $I_{Inst}$ on-shell implies that the $SO(3,C)$ gauge curvature $F^a_{\mu\nu}$ is Hodge self-dual with respect to a metric $g_{\mu\nu}$.  But $I_{Inst}$ also implies on-shell that $g_{\mu\nu}$ solves the Einstein equations, which in turn identifies $F^a_{\mu\nu}$ with the Riemann 
curvature $Riem\equiv{R}_{\mu\nu\rho\sigma}$.  Hence $Riem$ is also Hodge self-dual on any solution, which implies that the solutions of $I_{Inst}$ correspond to gravitational instantons.\footnote{The gauge curvature $F^a_{\mu\nu}$ takes its values in the Lie algebra $SO(3,C)$ corresponding to the self-dual half of the Lorentz group $SO(3,1)$.  The equivalence of internal self-duality with Hodge self-duality makes sense when one has a tetrad $\theta^I_{\mu}$, which intertwines between internal and spacetime indices.  But since tetrads are now derived quantities in $I_{Inst}$, this feature appears to be more fundamentally related to the Yang--Mills aspects of the theory.  We will show in a few sections that this is indeed the case.}

\section{Relation to the CDJ pure spin connection formulation}

There is an action for general relativity derived by Capovilla, Dell and Jacobson, which can be written almost entirely in terms of the spin connection \cite{SPINCON}.  The authors used Plebanski's action (\ref{STARTINGFROM}) as the starting point, from which they proceed to eliminate the 2-forms $\Sigma^a_{\mu\nu}$ and the matrix $\psi_{ae}$, leading for $\Lambda=0$ to the action 
\begin{eqnarray}
\label{CEEDEE}
I_{CDJ}=\int_Md^4x\hbox{tr}\bigl[M\bigl(M-{1 \over 2}(\hbox{tr}M)\bigr)\bigr],
\end{eqnarray}
\noindent
where we have defined
\begin{eqnarray}
\label{CEEDEE1}
M^{bf}=-{i \over {8\sqrt{-g}}}F^b_{\mu\nu}F^f_{\rho\sigma}\epsilon^{\mu\nu\rho\sigma}.
\end{eqnarray}
\noindent
Note that equation (\ref{CEEDEE1}) is the same as (\ref{OPTION761}), which is the symmetric part of (\ref{DECOMPOSE7}).  The action (\ref{OPTION78N}) serves in \cite{SPINCON} as an intermediate step in obtaining the 
action (\ref{CEEDEE1}) from (\ref{STARTINGFROM}).\footnote{Note that $\Psi$ in the present paper, after the elimination of the shift vector $N^i$ is actually defined as $\Psi^{-1}$ in \cite{SPINCON}.}  But in our 
context, equation (\ref{OPTION78N}) follows from (\ref{ISGIVEN2}) after elimination of $\Psi_{[ae]}$ and $N^i$ through their equations of motion.\par
\indent
Given that the CDJ action essentially follows from (\ref{ISGIVEN2}) after elimination of $\Psi_{ae}$, then this implies that $\Psi_{ae}$ should satisfy equation (2.20b) of \cite{SPINCON} on any solution for $\Lambda=0$.  We will show this by following the same steps in \cite{SPINCON}.  To obtain $\Psi_{ae}$ in terms of $A^a_{\mu}$, one would need to take the square root of $M^{bf}$ in (\ref{CEEDEE1}).  This introduces various complications, which are circumvented in \cite{SPINCON} by using the characteristic equation for (a symmetric) $\Psi_{ae}$
\begin{eqnarray}
\label{CEEDEE2}
\Psi^{-3}-(\hbox{tr}\Psi^{-1})\Psi^{-2}+{1 \over 2}\bigl((\hbox{tr}\Psi^{-1})^2-\hbox{tr}\Psi^{-2}\bigr)\Psi^{-1}-(\hbox{det}\Psi^{-1})=0.
\end{eqnarray}
\noindent
One must then use $\Psi^{-1}\Psi^{-1}=M$ from (\ref{OPTION761}) as well as $\hbox{tr}\Psi^{-1}=-\Lambda$ from (\ref{HAMM}), which when substituted into (\ref{CEEDEE2}) yields the equation
\begin{eqnarray}
\label{CEEDEE3}
\bigl(M+{1 \over 2}(-\hbox{tr}M+\Lambda^2)\bigr)\Psi^{-1}=-\Lambda{M}+I\sqrt{\hbox{det}M}
\end{eqnarray}
\noindent
where $I$ is the unit 3 by 3 matrix.  Then assuming that the left hand side of (\ref{CEEDEE3}) is invertible, then we can solve for $\Psi_{(ae)}$ as
\begin{eqnarray}
\label{CEEDEE4}
\Psi_{(ae)}=\bigl(-\Lambda_{af}+\delta_{af}(\hbox{det}M)^{1/2}\bigr)^{-1}\bigl(M_{fe}+{1 \over 2}\delta_{fe}(\Lambda^2-\hbox{tr}M)\bigr).
\end{eqnarray}
\noindent
Then upon substitution of (\ref{CEEDEE4}) into (\ref{OPTION78N}) one obtains the CDJ action (\ref{CEEDEE1}) for $\Lambda=0$.  For $\Lambda\neq{0}$ one can expand (\ref{CEEDEE4}) in powers of $\Lambda$ using a geometric series, yielding
\begin{eqnarray}
\label{CEEDEE5}
\Psi_{ae}=-{1 \over \Lambda}\Bigl(\delta_{ae}-\Bigl({{\Lambda(\Lambda^2-\hbox{tr}M)} \over {2(\hbox{det}M)^{1/2}}}+1\Bigr)\bigl(\delta_{ae}-\Lambda{M}_{ae}(\hbox{det}M)^{-1/2}\bigr)^{-1}\Bigr).
\end{eqnarray}
\noindent
Then one obtains the analogue of equation (3.9) of \cite{SPINCON}, which we will not repeat here.\par
\indent  
Let us now comment on the differences between (\ref{ISGIVEN2}) and (\ref{OPTION78N}), namely equation (2.8) in \cite{SPINCON}.  Equation (\ref{OPTION78N}) can be obtained by elimination of the 2-forms $\Sigma^a_{\mu\nu}$ directly from (\ref{STARTINGFROM}).  Then the CDJ action (\ref{CEEDEE}) follows by further elimination of the field $\Psi_{ae}$.  But (\ref{ISGIVEN2}) is the result of eliminating only $\Sigma^a_{ij}$, the spatial part of the 2-forms, and preserving the temporal components $\Sigma^a_{0i}$ as well as $\Psi_{ae}$.\footnote{The exception to this is the time gauge $e^0_i=0$, from which (\ref{ISGIVEN2}) follows.  This has the effect of fixing the boost parameters corresponding to the local Lorentz frame.  Recall that (\ref{STARTINGFROM}) is based on the self-dual $SU(2)_{-}$ part of the Lorentz algebra, which leaves open the interpretation of the antiself dual part $SU(2)_{+}$.  Since 
only $SU(2)_{-}$ is needed in order to obtain GR, it could be that $e^0_i$ is somehow associated with $SU(2)_{+}$.  On a separate note, we have preserved the temporal 2-form components $\Sigma^a_{0i}$ in $I_{Inst}$, in order to preserve the freedom to implementing the initial value constraints at the classical and the quantum levels.  The realization of this object will be relegated as a future direction of research.}  By complete elimination the complete 2-form $\Sigma^a$ as in \cite{SPINCON}, one also eliminates the ability to implement the Hamiltonian and diffeomorphism constraints in (\ref{DECOMP3}).  These are necessary for the construction of the metric $g_{\mu\nu}$, which plays the dual roles of solving the Einstein equations and enforcing Hodge duality.  Additionally, in equation (2.8) in \cite{SPINCON} the matrix $\Psi$ does not have an antisymmetric part, wherereas $\Psi_{[ae]}$ was necessary in order to obtain (\ref{DECOMPOSE}) as well as the shift vector $N^i$.  These two concepts are a vital part of the Hodge duality condition (\ref{DECOMPOSE7}).\par
\indent  
The Einstein equations can be derived from (\ref{CEEDEE}), which is shown as equations (2.19a), (2.19b) and (2.20a) in \cite{SPINCON}.  But the statement that the metric (equation (2.2) and (2.4) in \cite{SPINCON}) arises as a solution to these Einstein equations appears to the best of the present author's knowledge to be a separate postulate not derivable directly from (\ref{CEEDEE}).  We will show explicitly in the present paper that this metric is the same one arising from the Hodge duality condition (\ref{DECOMPOSE}), and complete the missing link in this loop regarding the Einstein equations.\par

\section{The spacetime metric: revisited}

We have shown that the instanton representation $I_{Inst}$, on-shell, implies a Hodge self-duality condition for the SO(3,C) curvature $F^a_{\mu\nu}$ with respect to a spacetime metric $g_{\mu\nu}$ solving the Einstein equations which also follow from $I_{Inst}$.  All that is needed to construct the 3-metric $h_{ij}$ for this spacetime metric are the spatial connection $A^a_i$ and the CDJ matrix $\Psi_{ae}$ solving the initial value 
constraints (\ref{DECOMP3}).  The specification of the shift vector $N^i$ via $A^a_0\subset{A}^a_{\mu}=(A^a_0,A^a_i)$ combined with a lapse function $N$ then completes the construction of $g_{\mu\nu}$ via (\ref{DECOMP}).  We will see that $I_{Inst}$ provides an additional simple formula for constructing $g_{\mu\nu}$ via the concept of Hodge duality.  The Hodge 
self-duality condition (\ref{SELFISH7}) is given by
\begin{eqnarray}
\label{HODGEIT1}
\epsilon_{ijk}B^j_aN^k+i\underline{N}h_{ij}B^j_a=-F^a_{0i}.
\end{eqnarray}
\noindent
Multiplying (\ref{HODGEIT1}) by $(B^{-1})^a_m$ and relabelling indices $m\rightarrow{j}$ we obtain the relation
\begin{eqnarray}
\label{HODGEIT2}
\epsilon_{ijk}N^k+i\underline{N}h_{ij}=-F^a_{0i}(B^{-1})^a_j.
\end{eqnarray}
\noindent
Equation (\ref{HODGEIT2}) provides a prescription for writing the spacetime metric explicitly in terms of the connection as follows.\footnote{In other words, the physical degrees of freedom from the initial value constraint contained in (\ref{DECOMP3}) become absorbed into the definition of the 3-metric $h_{ij}$.}  The antisymmetric part of (\ref{HODGEIT2}) yields the shift vector
\begin{eqnarray}
\label{HODGEIT3}
N^k=-{1 \over 2}\epsilon^{kij}F^a_{0i}(B^{-1})^a_j,
\end{eqnarray}
\noindent
and the symmetric part yields the 3-metric up to a conformal factor
\begin{eqnarray}
\label{HODGEIT4}
i\underline{N}h_{ij}=-F^a_{0(i}(B^{-1})^a_{j)}\equiv-{c}_{(ij)},
\end{eqnarray}
\noindent
where we have defined $c_{ij}=F^a_{0i}(B^{-1})^a_j$.  The determinant of (\ref{HODGEIT4}) yields
\begin{eqnarray}
\label{HOODGEE}
-i{{N^3} \over {\sqrt{h}}}=-\hbox{det}c_{(ij)}\equiv-{c}\longrightarrow{i}\underline{N}={c \over {N^2}}.
\end{eqnarray}
\noindent
Substituting this relation back into (\ref{HODGEIT4}) enables us to solve for $h_{ij}$
\begin{eqnarray}
\label{HOODGE}
h_{ij}=-\Bigl({{N^2} \over c}\Bigr)c_{(ij)}.
\end{eqnarray}
Let us define the following densitized object $\underline{c}_{(ij)}=c^{-1}c_{(ij)}$.  Then the line element (\ref{DECOMP}) is can also be written as\footnote{More precisely, since (\ref{DECOMP}) as defined by (\ref{DECOMP3}) forms a subset of the line element defined by (\ref{HOODGE1}), then the equality of (\ref{DECOMP}) with (\ref{HODGE1}) must be regarded as a consistency condition.  Since (\ref{HODGEIT1}) contains a velocity $\dot{A}^a_i$ and (\ref{DECOMP3}) does not, then the interpretation must be that the equality between the metrics enforces the evolution time of initial data satisfying the constraints.}
\begin{eqnarray}
\label{HOODGE1}
ds^2=g_{\mu\nu}{dx^{\mu}}dx^{\nu}=-N^2\bigl(dt^2+\underline{c}_{(ij)}\omega^i\otimes\omega^j\bigr)
\end{eqnarray}
\noindent
where we have defined the one forms $\omega^i=dx^i-N^idt$, with $N^i$ given by equation (\ref{HODGEIT3}).  Starting out with a spacetime of Lorentzian(Euclidean) signature for the lapse function $N$ real(imaginary), we have obtained a line element (\ref{HOODGE1}).  This implies the following consistency conditions
\begin{eqnarray}
\label{CONSISTENCY}
\underline{c}_{ij}>0\longrightarrow{N}~imaginary\longrightarrow{Euclidean~signature};\nonumber\\
\underline{c}_{ij}<0\longrightarrow{N}~real\longrightarrow{Lorentzian~signature}.
\end{eqnarray}
The result is that every connection $A^a_{\mu}$ with nondegenerate magnetic field $B^i_a$, combined with a lapse function $N$ determines a spacetime 
metric $g_{\mu\nu}$ of signature given by (\ref{CONSISTENCY}) solving the Einstein equations.\par
\indent
An elegant formula was constructed by Urbantke, which determines the metric with respect to which a given $SU(2)$ Yang--Mills curvature is self-dual, self-dual in the spacetime sense.  The formula is given by \cite{URBANTKE}
\begin{eqnarray}
\label{URBANT}
\sqrt{-g}g_{\mu\nu}={4 \over 3}\eta{f}_{abc}F^a_{\mu\rho}F^b_{\alpha\beta}F^c_{\sigma\nu}\epsilon^{\rho\alpha\beta\sigma}.
\end{eqnarray}
\noindent
Since we are treating general relativity in analogy to Yang--Mills theory, it is relevant to perform a 3+1 decomposition 
of (\ref{URBANT}).  The result of this decomposition is given by (see Appendix B for the details of the derivation)
\begin{eqnarray}
\label{URBANT2}
g_{00}\propto\hbox{det}(F^a_{0i});~~g_{0k}\propto\epsilon_{klm}(F^{-1})^{0l}_cB^m_c;~~g_{ij}\propto{F}^a_{0(i}(B^{-1})^a_{j)},
\end{eqnarray}
\noindent
Comparison of (\ref{URBANT2}) with (\ref{HODGEIT3}) and (\ref{HODGEIT4}) reveals that on-shell, the instanton representation of Plebanski gravity reproduces the Urbantke metric purely from an action principle.  When the spatial part of the Urbantke metric is built from variables solving the constraints (\ref{DECOMP3}), then the Urbantke metric solves the Einstein equations by construction.

\subsection{Reality conditions}

Since the connection $A^a_{\mu}$ is allowed to be complex, then the line element (\ref{HOODGE1}) in general allows for complex metrics $g_{\mu\nu}$.  Real general relativity corresponds to the restriction of this to real-valued metrics, which implies certain conditions on $A^a_{\mu}$ such that $\underline{c}_{ij}$ be real-valued in (\ref{CONSISTENCY}).  So the imposition of reality conditions requires that the undensitized 
matrix $c_{ij}=F^a_{0i}(B^{-1})^a_j$ be either real or pure imaginary, which leads to two cases
\begin{eqnarray}
\label{CONSISTENCY1}
c_{(ij)}~real\longrightarrow{Euclidean~signature};\nonumber\\
c_{(ij)}~imaginary\longrightarrow{Lorentzian~signature}.
\end{eqnarray}
\noindent
We will see that (\ref{CONSISTENCY1}) places restrictions on the connection $A^a_{\mu}$ for a spacetime of fixed signature.  For a general $A^a_{\mu}$ satisfying the reality conditions, there is apparently no constraint fixing the signature of the spatial part of the metric $h_{ij}$.\footnote{Hence there is a caveat associated with the labels `Euclidean' and `Lorentzian' used in (\ref{CONSISTENCY1}).  The lapse function $N$ is freely specifiable, since it is not constrained by $A^a_{\mu}$.  But it is still conceivable in (\ref{CONSISTENCY1}) that different components of $\underline{c}_{(ij)}$ could have different signs based on the initial data of $A^a_{\mu}$.  If this were to be the case, then this would bring in the possibility of topology changes of spacetimes described by $I_{Inst}$.}\par
\indent
The metric is clearly real if one is restricted to connections having a real curvature $F^a_{\mu\nu}$.  When $F^a_{\mu\nu}$ is complex then we must impose reality conditions requiring $\underline{c}_{ij}$ to be real as 
in (\ref{CONSISTENCY}).  The symmetric part of this enforces reality of the 3-metric $h_{ij}$ and the antisymmetric part enforces reality of the shift vector $N^i$.  The lapse function $N$ must always be chosen to be either real or pure imaginary.  But there is apparently no requirement in either case that the signature of $g_{\mu\nu}$ be fixed by $N$.  Note that the reality of the metric is a completely separate issue from the signature of spacetime, which in either case may apparently change.  This is unlike the case in the Ashtekar variables, where for Euclidean signature one is restricted to real variables.\par
\indent
We will delineate the reality conditions for the event where the curvature $F^a_{\mu\nu}$ is complex.  Split the connection $A^a_{\mu}$ into the real and imaginary parts of its spatial and temporal parts
\begin{eqnarray}
\label{REALITY}
A^a_i=(\Gamma-iK)^a_i;~~A^a_0=(\eta-i\zeta)^a.
\end{eqnarray}
\noindent
Corresponding to this 3+1 split there is a 3+1 split of $F^a_{\mu\nu}$ into spatial and temporal parts.  The spatial part is the magnetic field $B^i_a$ given by
\begin{eqnarray}
\label{REALITY1}
B^i_a=(R-iT)^i_a,
\end{eqnarray}
\noindent
where we have defined 
\begin{eqnarray}
\label{REALITY2}
R^i_a=\epsilon^{ijk}\partial_j\Gamma^a_k+{1 \over 2}\epsilon^{ijk}f_{abc}\Gamma^b_j\Gamma^c_k-{1 \over 2}\epsilon^{ijk}f^{abc}K^b_jK^c_k;\nonumber\\
T^i_a=\epsilon^{ijk}D_jK^a_k=\epsilon^{ijk}\bigl(\partial_jK^a_k+f^{abc}\Gamma^b_jK^c_k\bigr).
\end{eqnarray}
\noindent
The quantity $T^i_a$ is the covariant derivative of $K^a_i$ using $\Gamma^a_i$ as a connection.  The temporal part of the curvature $F^a_{\mu\nu}$ is given by
\begin{eqnarray}
\label{REALITY3}
F^a_{0i}=(f-ig)^a_i,
\end{eqnarray}
\noindent
where we have defined
\begin{eqnarray}
\label{REALITY4}
f^a_i=\dot{\Gamma}^a_i-D_i\eta^a+f^{abc}K^b_i\zeta^c;\nonumber\\
g^a_i=D_0K^a_i-D_i\zeta^a.
\end{eqnarray}
\noindent
The operator $D_i$ is the covariant derivative with respect to $\Gamma^a_i$ as in the second line of (\ref{REALITY2}), and $D_0$ is given by
\begin{eqnarray}
\label{REALTIY5}
D_0K^a_i=\dot{K}^a_i+f^{abc}\eta^bK^a_c.
\end{eqnarray}
\noindent
For the general complex case, reality conditions require that $c_{ij}=(B^{-1})^a_iF^a_{0j}$ be either real or pure imaginary as in (\ref{CONSISTENCY1}).  It will be convenient to use the following matrix identity, suppresing the indices
\begin{eqnarray}
\label{REALITY6}
B^{-1}=(R-iT)^{-1}=(1+iRT)\bigl(1-(R^{-1}T)^2\bigr)^{-1}R^{-1},
\end{eqnarray}
\noindent
which splits the inverse of a complex matrix into its real and imaginary parts.  Then upon contraction of the internal indices, $c_{ij}$ is given by
\begin{eqnarray}
\label{REALITY7}
(f-ig)(R-iT)^{-1}
=\bigl(f+gR^{-1}T+i(-g+fR^{-1}T)\bigr)\bigl(1-(R^{-1}T)^2\bigr)R^{-1}.
\end{eqnarray}
\noindent
The last two matrices in (\ref{REALITY7}) are real and the first matrix is in general complex.  For Lorentzian signature spacetimes we must require the real part of the first matrix to be zero, and for Euclidean signature we must require the imaginary part to be zero.  This leads to the matrix equations
\begin{eqnarray}
\label{REALITY8}
Euclidean~signature:~g^{-1}f=-R^{-1}T;\nonumber\\
Lorentzian~signature:~f^{-1}g=R^{-1}T.
\end{eqnarray}
\noindent
The aforementioned caveats still apply with respect to the stability of the signature.  But in either case the reality conditions constitute 9 equations in 24 unknowns, namely the 12 complex components of 
the 4-dimensional SO(3,C) connection $A^a_{\mu}$.  After implementation of these reality conditions, then this leaves $24-9=15$ real degrees of freedom in $A^a_{\mu}$.\footnote{For example in (\ref{REALITY}), then one possibility is to regard the nine components of $\Re\{A^a_i\}$ as freely specifiable, and then use (\ref{REALITY8}) to determine the nine components of $\Im\{A^a_i\}$ in terms of them.}

\section{Gravity as a `generalized' Yang--Mills theory}

We will now show how the concept of Hodge self-duality stems at a more fundamental level from internal duality with respect to gravitational degrees of freedom.  Let us start off by considering the following action which 
resembles $SO(3)$ Yang--Mills theory in curved space time
\begin{eqnarray}
\label{MEALACTION}
I=\int_Md^4x\Bigl[-{1 \over 4}\sqrt{-g}g^{\mu\nu}g^{\rho\sigma}F^b_{\mu\nu}F^f_{\rho\sigma}\Psi_{bf}+{1 \over G}\sqrt{-g}R[g]\Bigr].
\end{eqnarray}
\noindent
The quantity $g^{\mu\nu}$ is the covariant metric corresponding to the background spacetime upon which a Yang--Mills field $A^a_{\mu}$ propagates, and the Yang--Mills curvatures in (\ref{MEALACTION}) 
and $F_{\mu\nu}$ is the curvature of $A^a_{\mu}$, given by
\begin{eqnarray}
\label{MEALACTION1}
F^a_{\mu\nu}=\partial_{\mu}A^a_{\nu}-\partial_{\nu}A^a_{\mu}+f^{abc}A^b_{\mu}A^c_{\nu}
\end{eqnarray}
\noindent
where $f^{abc}=\epsilon^{abc}$ are the structure constants of $SO(3)$.  Equation (\ref{MEALACTION}) is different from the usual Yang--Mills theory in that the two curvatures $F^a_{\mu\nu}$ additionally couple to a field $\Psi_{bf}$ taking its values in two copies of $SO(3)$.   In the special case $\Psi_{ae}=k\delta_{ae}$ for some numerical constant $k$, plays the $\Psi_{ae}$ role of the Cartan--Killing metric for the $SO(3)$ gauge group.  There is a wide array of literature concerning gravity and Yang--Mills theory, where one attempts to solve (\ref{MEALACTION}) for the Yang--Mills field $A^a_{\mu}$ as well as for the metric $g_{\mu\nu}$.  But in the gravitational 
context, $\Psi_{ae}=-{3 \over \Lambda}\delta_{ae}$ implies that the metric $g_{\mu\nu}$ must be restricted to spacetimes of Petrov Type O, since $\Psi_{ae}$ then has three equal eigenvalues \cite{PENROSERIND}.\par
\indent
The implication is that when one solves (\ref{MEALACTION}) in the case $\Psi_{ae}=-{3 \over \Lambda}\delta_{ae}$, then one is solving the coupled Yang--Mills only for conformally flat spacetimes.  But we would like to incorporate more general geometries.  On the one hand in vacuum Yang--Mills theory one already has a Yang--Mills solution for known metrics by virtue of Hodge duality and the Bianchi identity.  On the other hand, the generalization of $\Psi_{ae}$ to include gravitational degrees of freedom, as we will see, enables one to identify the Yangs--Mills with the gravity theory that it is coupling to.  To see this, let us split the Yang--Mills part of the Lagrangian of (\ref{MEALACTION}) into its spatial and temporal parts
\begin{eqnarray}
\label{MEALACTION2}
L_{YM}={{\sqrt{-g}} \over 2}\Bigl(g^{00}g^{ij}F^b_{0i}F^f_{0j}-g^{0i}g^{0j}F^b_{0i}F^f_{0j}+2g^{0i}g^{jk}F^b_{ij}F^f_{0k}+{1 \over 2}g^{ik}g^{jl}F^b_{ij}F^f_{kl}\Bigr)\Psi_{bf},
\end{eqnarray}
\noindent
where $F^a_{0i}=\dot{A}^a_i-D_iA^a_0$ is the temporal component of the curvature.  The electric field is the momentum canonically conjugate to the connection
\begin{eqnarray}
\label{MEALACTION3}
\Pi^i_b={{\delta{I}_{YM}} \over {\delta\dot{A}^b_i}}=\sqrt{-g}\bigl(g^{00}g^{ij}F^f_{0j}-g^{0i}g^{0j}F^f_{0j}+g^{0m}g^{ni}F^f_{mn}\bigr)\Psi_{bf}.
\end{eqnarray}
\noindent
Next, we will make use of the 3+1 decomposition of the spacetime metric
\begin{displaymath}
g^{\mu\nu}=
\left(\begin{array}{cc}
g^{00} & g^{0i}\\
g^{0j} & g^{ij}\\
\end{array}\right)
=
\left(\begin{array}{cc}
-{1 \over {N^2}} & {{N^i} \over {N^2}}\\
{{N^j} \over {N^2}} & h^{ij}-{{N^iN^j} \over {N^2}}\\
\end{array}\right)
,
\end{displaymath}
\noindent
where $N^{\mu}=(N,N^i)$ are the lapse function and shift vector, and $\sqrt{-g}=N\sqrt{h}$ is the determinant of $g_{\mu\nu}$.  Substitution into (\ref{MEALACTION3}) yields
\begin{eqnarray}
\label{MEALACTION4}
\Pi^i_b={{\sqrt{h}} \over N}\bigl(-h^{ij}F^f_{0j}+N^mh^{ni}F^f_{mn}\bigr)\Psi_{bf},
\end{eqnarray}
\noindent
and substitution into (\ref{MEALACTION2}) yields
\begin{eqnarray}
\label{MEALACTION5}
L_{YM}=-{1 \over 2}N\sqrt{h}\Bigl(-{1 \over {N^2}}h^{ij}F^b_{0i}F^f_{0j}+2{{N^i} \over {N^2}}\Bigl(h^{jk}-{{N^jN^k} \over {N^2}}\Bigr)F^b_{ij}F^f_{0k}\nonumber\\
+{1 \over 2}h^{ik}\Bigl(h^{jl}-{{2N^jN^l} \over {N^2}}\Bigr)F^b_{ij}F^f_{kl}\Bigr)\Psi_{bf}.
\end{eqnarray}
\noindent
We will now eliminate the velocities $\dot{A}^a_i$ from (\ref{MEALACTION5}) by inverting (\ref{MEALACTION4})
\begin{eqnarray}
\label{MEALACTION6}
F^f_{0j}=h_{jk}\Bigl(-{N \over {\sqrt{h}}}\Pi^k_b(\Psi^{-1})^{bf}+N^mh^{nk}F^f_{mn}\Bigr).
\end{eqnarray}
\noindent
Upon substitution of (\ref{MEALACTION6}) into (\ref{MEALACTION5}) after several long but straightforward algebraic steps, we obtain
\begin{eqnarray}
\label{MEALACTION7}
L_{YM}={1 \over 2}{N \over {\sqrt{h}}}h_{ij}\Pi^i_b\Pi^j_f(\Psi^{-1})^{bf}+{1 \over 4}N\sqrt{h}h^{ik}h^{jl}F^b_{ij}F^f_{kl}\Psi_{bf}.
\end{eqnarray}
\noindent
Defining the $SO(3,C)$ magnetic field by $F^a_{ij}=\epsilon_{ijk}B^k_a$, and using the relation
\begin{eqnarray}
\label{MEALACTION8}
{1 \over 2}\epsilon_{ijm}\epsilon_{kln}h^{ik}h^{jl}={1 \over h}h_{mn},
\end{eqnarray}
\noindent
and presupposing the 3-metric $h_{ij}$ to be nondegenerate, then (\ref{MEALACTION7}) yields 
\begin{eqnarray}
\label{MEALACTION9}
L_{YM}={1 \over 2}\underline{N}h_{ij}\bigl((\Psi^{-1})^{bf}\Pi^i_b\Pi^j_f-\Psi_{bf}B^i_bB^j_f\bigr).
\end{eqnarray}
\noindent
This is the electromagnetic decomposition of the generalized Yang--Mills action, with $\Psi_{bf}$ replacing the Cartan--Killing form for the gauge group.\par
\indent
To see how GR follows from this Yang--Mills theory, let us impose the following relation between the electric and the magnetic fields of the former
\begin{eqnarray}
\label{CANB}
\Pi^i_a=\beta\Psi_{ae}B^i_e
\end{eqnarray}
\noindent  
for some numerical constant $\beta$.  Then for nondegenerate $\Psi_{bf}$, substitution of (\ref{CANB}) into (\ref{MEALACTION3}) implies that
\begin{eqnarray}
\label{CANBE}
\beta{B}^i_f=N\sqrt{h}\bigl(g^{00}g^{ij}F^f_{0j}-g^{0i}g^{0j}F^f_{0j}+g^{0m}g^{ni}F^f_{mn}\bigr).
\end{eqnarray}
\noindent
The right hand side of (\ref{CANBE}) is given by
\begin{eqnarray}
\label{CANBE1}
N\sqrt{h}\Bigl[-{1 \over {N^2}}\Bigl(h^{ij}-{{N^iN^j} \over {N^2}}\Bigr)F^f_{0j}-{{N^iN^j} \over {N^4}}F^f_{0j}-{{N^m} \over {N^2}}\Bigl(h^{ni}-{{N^nN^i} \over {N^2}}\Bigr)F^f_{mn}\Bigr)\Bigr]
\end{eqnarray}
\noindent
which simplifies to
\begin{eqnarray}
\label{CANBE2}
{{\sqrt{h}} \over N}\bigl(h^{ij}F^f_{0j}+N^kh^{ij}F^f_{kj}\bigr)=-\beta{B}^i_f.
\end{eqnarray}
\noindent
Equation (\ref{CANBE2}) can be rewritten as
\begin{eqnarray}
\label{CANBE3}
F^f_{0j}+\epsilon_{jmk}B^m_gN^k+\beta\underline{N}h_{ji}B^i_f=0.
\end{eqnarray}
\noindent
The choice $\beta=\pm{i}$ would imply that equation (\ref{CANB}) automatically imposes Hodge self-duality of the Yang--Mills curvature $F^f_{\mu\nu}$ with respect to the metric $g_{\mu\nu}$ which it couples to, namely 
\begin{eqnarray}
\label{CANBE4}
H^{\mu\nu\rho\sigma}F^b_{\rho\sigma}=0,
\end{eqnarray}
\noindent
where we have defined the Hodge self-duality operator
\begin{eqnarray}
\label{CANBE5}
H^{\mu\nu\rho\sigma}={1 \over 2}\Bigl(\sqrt{-g}\bigl(g^{\mu\rho}g^{\nu\sigma}-g^{\nu\rho}g^{\mu\sigma}\bigr)+\beta\epsilon^{\mu\nu\rho\sigma}\Bigr).
\end{eqnarray}
\noindent
Comparison of (\ref{CANB}) with the spatial restriction of equation the third equation of (\ref{STARTINGFROM3}), and comparison of (\ref{CANBE3}) with (\ref{DECOMPOSE7}), implies that (\ref{CANB}) is the internal analogue of Hodge self-duality.  Indeed, the fact that the metric defining (\ref{CANBE4}) solves the Einstein equations transforms (\ref{OPTION78N}) on-shell into (\ref{MEALACTION}).  Since the solutions to ordinary vacuum Yang--Mills theory include Yang--Mills instantons, then this suggests that $I_{Inst}$ is a theory which should include gravitational instantons.

\section{Gravitational instantons: Revisited}

We will now put into context the points raised in the introduction section regarding the apparent ambiguity in the definition of gravitational instantons.  It has been noted by Ashtekar and Renteln \cite{ASH1} that the Ansatz
\begin{eqnarray}
\label{THATTHE}
B^i_a=-{\Lambda \over 3}\widetilde{\sigma}^i_a,
\end{eqnarray}
\noindent
solves the initial value constraints of the Ashtekar variables arising from (\ref{ASHTESH}).  It was noted that this corresponds to the conformally flat spacetimes.\footnote{We will see that (\ref{CANB}) is the 
generalization of (\ref{THATTHE}) which incorporates more general geometries including Types D and O, when $\Psi_{ae}$ becomes identified with the $\Psi_{ae}$ of $I_{Inst}$.}  There is a covariant form of the action (\ref{ASHTESH}) provided by Samuel \cite{SAMUEL}, \cite{SAMUEL1}) in which the basic variables are two forms $\Sigma^b={1 \over 2}\Sigma^b_{\mu\nu}{dx^{\mu}}\wedge{dx^{\nu}}$, given by
\begin{eqnarray}
\label{THATTHE1}
I=\int_Md^4x\bigl(\Sigma^b_{\mu\nu}F^b_{\rho\sigma}+{\Lambda \over 6}\Sigma^b_{\mu\nu}\Sigma^b_{\rho\sigma}\bigr)\epsilon^{\mu\nu\rho\sigma}.
\end{eqnarray}
\noindent
Equation (\ref{THATTHE1}) leads to general relativity with cosmological constant through the equations of motion
\begin{eqnarray}
\label{THATTHE2}
\epsilon^{\mu\nu\rho\sigma}D_{\nu}\Sigma^b_{\rho\sigma}=0;~~F^b_{\mu\nu}=-{\Lambda \over 3}\Sigma^b_{\mu\nu},
\end{eqnarray}
\noindent
where the two form is constructed from SL(2,C) one forms 
\begin{eqnarray}
\label{THATTHE21}
\Sigma^{AB}_{\mu\nu}=i\bigl(\theta_{\mu}^{AA^{\prime}}\theta_{\nu{A}^{\prime}}^B-\theta_{\nu}^{AA^{\prime}}\theta_{\mu{A}^{\prime}}^B\bigr)
\end{eqnarray}
\noindent
in self-dual combination.  The class of solutions described by the second equation of (\ref{THATTHE2}) are the evolution of (\ref{THATTHE}), which is the data set on the initial spatial hypersurface.  The observation that the first equation of (\ref{THATTHE2}) follows identically from the second due to the Bianchi identity, combined with the self duality in (\ref{THATTHE21}) allows an association of gravity to Yang--Mills instantons to be inferred \cite{SAMUEL}.\par
\indent
It was postulated that there might be other Yang--Mills field strengths which satisfy (\ref{THATTHE2}), but one is limited to conformally flat metrics since not all two forms $\Sigma$ are constructible from 
tetrad one forms $\theta_{\mu}$ as in (\ref{THATTHE21}).  The problem of relating (\ref{THATTHE2}) to the Yang--Mills self-duality condition $^{*}F=F$ resides in the observation that the metric $g_{\mu\nu}$ must first be 
known.  In \cite{GRAVINTGAUGE}, Jacobson eliminates the tetrad from the self-duality condition to address the sector with vanishing self-dual Weyl curvature, by proposing the following condition on the curvature
\begin{eqnarray}
\label{THATTHE4}
{F^b}\wedge{F^f}-{1 \over 3}\delta^{bf}\hbox{tr}{F}\wedge{F}=0.
\end{eqnarray}
\noindent
Given a connection $A^a_{\mu}$ which solves (\ref{THATTHE4}), the tetrads in (\ref{THATTHE21}) associated with the 2-form $\Sigma^b$ determine a metric which is a self-dual Einstein solution with cosmological constant $\Lambda$.  Moreover, the curvature satisfying (\ref{THATTHE4}) is self-dual with respect to this metric.  Since (\ref{THATTHE4}) is the same as the second equation of (\ref{STARTINGFROM3}) when $\Psi_{ae}\propto\delta_{ae}$, then the problem of `finding the metric' as pointed out by Samuel in \cite{SAMUEL} translates into the problem of finding the connection in (\ref{THATTHE4}).\par
\indent
Hence the aforementioned developments have been shown only for the conformally self-dual case where the self-dual Weyl tensor $\psi_{ae}$ vanishes, whence the metric is explicitly constructible.  This limits one to spacetimes of Petrov Type O.\footnote{In \cite{CAPBLANSKI} gravitational instantons are defined as space-times with vanishing self-dual Weyl curvature, and nonvanishing cosmological constant.  This falls within the 
Petrov Type O with $\Psi_{ae}=-{3 \over \Lambda}\delta_{ae}$, with no restrictions on the connection $A^a_i$.  We would like to generalize this to incorporate Type D and I space-times.}  The proposition of the present paper has been to extend the library of solutions to include the Petrov Type I and D cases using $I_{Inst}$. 

\subsection{Generalization beyond Petrov Type O instantons}
We have seen that the CDJ Ansatz, the spatial restriction of the third equation of (\ref{STARTINGFROM3}), imposes the condition of Hodge self-duality on the `generalized' $SO(3)$ Yang--Mills fields in (\ref{CANB}).  When $\Psi_{ae}$ is chosen to satisfy the constraints (\ref{DECOMP3}), then the implication is that this Yang--Mills theory becomes a theory of GR.  Since vacuum Yang--Mills theory in conformally flat spacetimes describes instantons, then this suggests that the gravitational analogue of pure Yang--Mills theory must describe gravitational instantons, specifically incorporating the physical degrees of freedom from (\ref{DECOMP3}).  To examine the implications for gravity let us recount 
the action (\ref{OPTION78N}), repeated here for completeness
\begin{eqnarray}
\label{GRAVINT}
I_{Inst}=\int_Md^4x{1 \over 8}\Psi_{ae}F^a_{\mu\nu}F^e_{\rho\sigma}\epsilon^{\mu\nu\rho\sigma}-i\sqrt{-g}\bigl(\Lambda+\hbox{tr}\Psi^{-1}\bigr),
\end{eqnarray}
which corresponds to (\ref{OPTION78}) at the level after elimination of the shift vector $N^i$.  Recall also that the equation of motion for $\Psi_{ae}$ prior to elimination of $N^i$ and in (\ref{DECOMPOSE}) implies the Hodge self-duality condition
\begin{eqnarray}
\label{GRAVINT1}
\beta\epsilon^{\mu\nu\rho\sigma}F^a_{\mu\nu}=\sqrt{-g}g^{\mu\rho}g^{\nu\sigma}F^a_{\rho\sigma}.
\end{eqnarray}
\noindent
once one has made the identification of $h_{ij}=h_{ij}[\Psi,A]$.  Substitution of (\ref{GRAVINT1}) into the first term of (\ref{GRAVINT}) yields
\begin{eqnarray}
\label{GRAVINT2}
I_{Inst}=\int_Md^4x{\beta \over 4}\sqrt{-g}g^{\mu\rho}g^{\nu\sigma}F^a_{\mu\nu}F^e_{\rho\sigma}\Psi_{ae}-i\sqrt{-g}\bigl(\Lambda+\hbox{tr}\Psi^{-1}\bigr),
\end{eqnarray}
\noindent
which is nothing other than the action for gravity coupled to a `generalized' $SO(3)$ Yang--Mills theory of gravity (\ref{MEALACTION}).  On the other hand, the equation of motion for $\Psi_{ae}$ derived from (\ref{GRAVINT}) is
\begin{eqnarray}
\label{SHOODGEET}
{1 \over 8}F^b_{\mu\nu}F^f_{\rho\sigma}\epsilon^{\mu\nu\rho\sigma}+i\sqrt{-g}(\Psi^{-1}\Psi^{-1})^{fb}=0.
\end{eqnarray}
\noindent
Comparison of (\ref{SHOODGEET}) with (\ref{METRIC}) indicates that dynamically on the solution to the equations of motion,
\begin{eqnarray}
\label{ELIMINATE11}
{1 \over 8}F^b_{\mu\nu}F^f_{\rho\sigma}\epsilon^{\mu\nu\rho\sigma}=-i\beta^{-1/2}N(\hbox{det}B)^{-1/2}(\hbox{det}\Psi)^{-1/2}h_{ij}B^i_bB^j_f=
-i\beta\underline{N}h_{ij}B^i_bB^j_f
\end{eqnarray}
\noindent
where $\underline{N}=Nh^{-1/2}$.  Since the initial value constraints must be consistent with the equations of motion we can insert (\ref{ELIMINATE11}) into (\ref{GRAVINT}), which yields 
\begin{eqnarray}
\label{RELATION3}
I_{Inst}={\beta \over 2}\int_M\Psi_{ae}{F^a}\wedge{F^e}=-i\beta\int_M\underline{N}h_{ij}\Psi_{ae}B^i_aB^j_ed^4x.
\end{eqnarray}
\noindent
But equation (\ref{RELATION3}) is only the magnetic part of a Yang--Mills theory in curved spacetime.  To obtain the electric part we use 
the relation $B^i_e={1 \over \beta}\Psi^{-1}_{ae}\widetilde{\sigma}^i_a$, which shows on-shell that the following objects are equivalent
\begin{eqnarray}
\label{RAOULT}
-i\beta\underline{N}h_{ij}B^i_bB^j_f\Psi_{bf}=-i\underline{N}h_{ij}\widetilde{\sigma}^i_bB^j_f=-i\beta\underline{N}h_{ij}(\Psi^{-1})^{bf}\widetilde{\sigma}^i_b\widetilde{\sigma}^j_f.
\end{eqnarray}
\noindent
So we can use (\ref{RAOULT}) to eliminate $B^i_a$ from (\ref{RELATION3}), yielding
\begin{eqnarray}
\label{RELATION4}
I_{Inst}={\beta \over 2}\int_M\Psi_{ae}{F^a}\wedge{F^e}=-i\beta\int_M{1 \over {\beta^2}}\underline{N}h_{ij}(\Psi^{-1})^{ea}\widetilde{\sigma}^i_a\widetilde{\sigma}^j_ed^4x.
\end{eqnarray}
\noindent
The action for GR in the instanton representation evaluated on a classical solution can be written as the average of
(\ref{RELATION3}) and (\ref{RELATION4}), which yields
\begin{eqnarray}
\label{RELATION5}
I_{Inst}={i \over 2}\int{dt}\int_{\Sigma}d^3x\underline{N}h_{ij}\Bigl[-{1 \over {\beta^2}}(\Psi^{-1})^{bf}\widetilde{\sigma}^i_b\widetilde{\sigma}^j_f-\Psi_{bf}B^i_bB^j_f\Bigr]\nonumber\\
=i\beta\int{dt}\int_{\Sigma}d^3x\Bigl[\underline{N}h_{ij}T^{ij}-{i \over 2}\beta\Bigl(1+{1 \over {\beta^2}}\Bigr)\underline{N}h_{ij}\widetilde{\sigma}^i_b\widetilde{\sigma}^j_f(\Psi^{-1})^{bf}\Bigr]
\end{eqnarray}
\noindent
with $T^{ij}$ given by
\begin{eqnarray}
\label{RELATION6}
T^{ij}={1 \over 2}\bigl((\Psi^{-1})^{ae}\widetilde{\sigma}^i_a\widetilde{\sigma}^j_e-\Psi_{ae}B^i_aB^j_e\bigr).
\end{eqnarray}
\noindent
With the exception of the term proportional to $\beta$, (\ref{RELATION5}) would be the action for a `generalized' Yang--Mills theory.  Note that it is a genuine Yang--Mills theory only for $\Psi_{ae}=k\delta_{ae}$, which covers only the Type O sector of gravity.  Upon making the identification $\widetilde{\sigma}^i_a\equiv\Pi^i_a$ from (\ref{MEALACTION9}), then we have on the solution to the equations of motion that
\begin{eqnarray}
\label{RELATION7}
{1 \over 8}\int_Md^4x\Psi_{bf}F^b_{\mu\nu}F^f_{\rho\sigma}\epsilon^{\mu\nu\rho\sigma}=i\beta\int_Md^4x\sqrt{-g}g^{\mu\rho}g^{\nu\sigma}F^b_{\mu\nu}F^f_{\rho\sigma}+Q,
\end{eqnarray}
\noindent
where $Q$ is the bottom line of (\ref{RELATION5}).  The identification between the Yang--Mills and the instanton representation actions can be made only for $\beta^2=-1$.  In this case $Q=0$ and equation (\ref{RELATION7}) implies on the solution to the equations of motion that
\begin{eqnarray}
\label{RELATION8}
{1 \over 8}\int_Md^4x\Bigl(\sqrt{-g}\bigl(g^{\mu\rho}g^{\nu\sigma}-g^{\nu\rho}g^{\mu\sigma}\pm{i}\epsilon^{\mu\nu\rho\sigma}\Bigr)F^b_{\mu\nu}F^f_{\rho\sigma}\Psi_{bf}=0.
\end{eqnarray}
\noindent
In order for this to be true for all curvatures, we must have 
\begin{eqnarray}
\label{RELATION9}
\pm{i \over 2}\epsilon^{\mu\nu\rho\sigma}F^f_{\rho\sigma}=\sqrt{-g}g^{\mu\rho}g^{\nu\sigma}F^f_{\rho\sigma},
\end{eqnarray}
\noindent
namely that the curvature of the starting theory must be self-dual in the Hodge sense in any solution to the equations of motion.  In this case, it can be said that general relativity is literally a Yang--Mills theory coupled gravitationally to itself.\par
\indent

\section{Summary}

In this paper we have presented the instanton representation of Plebanski gravity, a new formulation of general relativity.  The basic dynamical variables 
are an $SO(3,C)$ gauge connection $A^a_{\mu}$ and a matrix $\Psi_{ae}$ taking its values in two copies of $SO(3,C)$.  The consequences of the associated 
action $I_{Inst}$ were determined via its equations of motion with the following results (i) The two equations of motion for $I_{Inst}$ imply the Einstein equations when the initial value constraints are satisfied. (ii) When these constraints are satisfied, then one can define a spatial 3-metric $h_{ij}[\Psi,A]$ using $\Psi_{ae}$ and $A^a_i$, the spatial part of the connection $A^a_{\mu}$.  (iii) The first equation of motion 
for $I_{Inst}$ is consistent with the second equation when the intial value constraints are satisfied. (iv) The first equation of motion of $I_{Inst}$ implies that the curvature $F^a_{\mu\nu}$ is Hodge self-dual with respect to the metric $g_{\mu\nu}$ which solves the Einstein equations as a consequence of the initial value constraints. Each of these results hinges crucially on the existence of solutions to the initial value constraints.  So it remains to be verified that that once the initial value constraints are satisfied on an initial spatial hypersurface, then the equations of motion should preserve these constraints for all time.  We have relegated the demonstration of 
this to \cite{EYOITA1}.\par
\indent
Additionally, we have clarified the relation between $I_{Inst}$ and $I_{CDJ}$ in \cite{CAP}.  The two formulations are not the same as it may naively appear for the following reasons.  (i) The action $I_{CDJ}$ at the level prior to elimination of $\Psi_{ae}$ from $I_{Pleb}$ is missing the 2-forms $\Sigma^a_{\mu\nu}$ as well as the antisymmetric part of $\Psi_{ae}$.  However, $I_{Inst}$ contains $\Sigma^a_{0i}$, the temporal part of $\Sigma^a_{\mu\nu}$ as well as $\Psi_{[ae]}$. (ii) The Hodge duality condition follows directly as an equation of motion for $I_{Inst}$, a crucial part of which involves $N^{\mu}=(N,N^i)$ from $\Sigma^a_{0i}$ which are needed both for consructing GR solutions as well as for implementing the initial value constraints.\footnote{The advantages of these features should become more apparent when one proceeds to construct GR solutions and in the quantum theory.} (iii) The reality conditions 
in $I_{Inst}$ appear to be intimately connected with the signature of spacetime as well as initial data, which is unlike the usual GR formulations.  The implications of this should be borne out when one attempts to construct solutions.\par
\indent
The instanton representation $I_{Inst}$ has exposed an interesting relation between general relativity and Yang--Mills theory, which suggests that this is indeed a theory of `generalized' Yang--Mills instantons.  In the conformally flat case, the CDJ matrix $\Psi_{ae}$ has three equal eigenvalues and thus plays the role of a Cartan--Killing $SO(3)$ invariant metric.  The generalization of this to more general geometries presents an interesting physical interpretation, since $\Psi_{ae}$ contains the true gravitational degrees of freedom.  In the Petrov Type D case for example, where $\Psi_{ae}$ has two equal eigenvalues, then the Yang--Mills SO(3) symmetry is broken to $SO(2)$.  In the algebraically general Type I case, where $\lambda_1\neq\lambda_2\neq\lambda_3$, the $SO(3,C)$ symmetry is completely broken.  A possible future direction is to investigate possible mechanisms which could induce such a breaking of this symmetry.\par
\indent
Nevetheless, the first order of business in future research will be to check for consistency of the initial value constraints of $I_{Inst}$ under time evolution.  Then next will be to use $I_{Inst}$ reconstruct as many of the known GR solutions as possible and to construct new solutions.  Additionally, we will examine the quantum theory with a view to addressing many of the questions in quantum gravity.
\subsection{Preview into the quantum theory}
Instantons in Yang--Mills theory can be associated with transitions between topologically inequivalent vacua, induced by tunneling classical solutions upon Wick rotation between Lorentzian and Euclidean signature spacetimes.  A future direction of research will be to investigate the analouge of this feature within $I_{Inst}$, in addition to the quantum aspects of the theory.  For instance, upon substitution of 
contraction of (\ref{SHOODGEET}) with $\Psi_{bf}$ one obtains
\begin{eqnarray}
\label{SHOODGEET1}
{1 \over 8}\Psi_{bf}F^b_{\mu\nu}F^f_{\rho\sigma}\epsilon^{\mu\nu\rho\sigma}=-i\sqrt{-g}\hbox{tr}\Psi^{-1}=i\sqrt{-g}\Lambda,
\end{eqnarray}
\noindent
where we have used the Hamiltonian constraint from variation of $N$ in (\ref{GRAVINT}).  Substitution of (\ref{SHOODGEET1}) back into (\ref{GRAVINT}) yields
\begin{eqnarray}
\label{SHOODGEET2}
I_{Inst}=i\Lambda\int_Md^x\sqrt{-g}=i\Lambda{Vol}(M),
\end{eqnarray}
\noindent
where $Vol(M)$ is the spacetime volume.  The exponentiation of this in units of $\hbar{G}$ yields
\begin{eqnarray}
\label{SHOODGEET3}
\boldsymbol{\psi}=e^{i\Lambda(\hbar{G})^{-1}Vol(M)},
\end{eqnarray}
\noindent
which forms the dominant contribution to the path integral for gravity due to gravitational instantons \cite{CARLIP}.  On the other hand, substitution of $\Psi_{ae}=-{3 \over \Lambda}\delta_{ae}$ into the starting 
action (\ref{OPTION78}) produces a total derivative leading via Stokes' theorem to a Chern--Simons boundary term $I_{CS}$.  The exponentiation of this boundary term in units of $\hbar{G}$ yields
\begin{eqnarray}
\label{KOODAAMAA}
\boldsymbol{\psi}_{Kod}=e^{\pm{3}\hbar{G}\Lambda{I}_{CS}[A]},
\end{eqnarray}
\noindent
which is known as the Kodama state which describes DeSitter spacetime \cite{KOD}, \cite{POSLAMB}.  Part of the quantum theory of $I_{Inst}$ will be to clarify the role of (\ref{KOODAAMAA}) in quantum gravity, and to attempt to find its counterparts for $\Psi_{ae}$ corresponding to more general spacetime geometries.

\section{Appendix A: Components of the Hodge self-duality operator}

From the equation
\begin{eqnarray}
\label{FROMTHE}
N\sqrt{h}\Bigl(\bigl(g^{00}g^{kj}-g^{k0}g^{0j}\bigr)F^a_{0j}+g^{0i}g^{kj}\epsilon_{ijm}B^m_a\Bigr)=\beta{B}^k_a
\end{eqnarray}
\noindent
from (\ref{SELFISH3}), we have
\begin{eqnarray}
\label{FROMTHE1}
N\sqrt{h}\Bigl(\Bigl(-{1 \over {N^2}}\Bigr)\Bigl(h^{kj}-{{N^kN^j} \over {N^2}}\Bigr)-\Bigl({{N^kN^j} \over {N^2}}\Bigr)\Bigr)F^a_{0j}\nonumber\\
+N\sqrt{h}\Bigl(-{{N^i} \over {N^2}}\Bigr)\Bigl(h^{kj}-{{N^kN^j} \over {N^2}}\Bigr)\epsilon_{ijm}B^m_a\Bigr)=\beta{B}^k_a.
\end{eqnarray}
\noindent
Cancelling off the terms multipying $F^a_{0j}$ which are quadratic in $N^i$, we have
\begin{eqnarray}
\label{FROMTHE2}
-\Bigl({{\sqrt{h}} \over N}\Bigr)h^{kj}\bigl(F^a_{0j}+\epsilon_{jmi}B^m_aN^i\bigr)=\beta{B}^k_a.
\end{eqnarray}
\noindent
Multiplying (\ref{FROMTHE2}) by $\underline{N}=N/\sqrt{h}$ and by $h_{lk}$, this yields
\begin{eqnarray}
\label{FROMTHE3}
F^a_{0l}+\epsilon_{lmi}B^m_aN^i+\beta\underline{N}h_{lk}B^k_a=0.
\end{eqnarray}
\noindent
From the equation
\begin{eqnarray}
\label{FROMTHE4}
N\sqrt{h}\Bigl(\bigl(g^{m0}g^{nj}-g^{n0}g^{mj}\bigr)F^a_{0j}+g^{mi}g^{nj}\epsilon_{ijk}B^k_a\Bigr)=\beta\epsilon^{mn0j}F^a_{0j},
\end{eqnarray}
\noindent
from (\ref{SELFISH5}), we have
\begin{eqnarray}
\label{FROMTHE5}
N\sqrt{h}\Bigl[-\Bigl({{N^m} \over {N^2}}\Bigr)\Bigl(h^{nj}-{{N^nN^j} \over {N^2}}\Bigr)+\Bigl({{N^n} \over {N^2}}\Bigr)\Bigl(h^{mj}-{{N^mN^j} \over {N^2}}\Bigr)\Bigr)F^a_{0j}\nonumber\\
+\Bigl(h^{mi}-{{N^mN^i} \over {N^2}}\Bigr)\Bigl(h^{nj}-{{N^nN^j} \over {N^2}}\Bigr)\epsilon_{ijk}B^k_a\Bigr]=\beta\epsilon^{0mnj}F^a_{0j}.
\end{eqnarray}
\noindent
expanding and using the vanishing of the term quatic in the shift vecto $N^i$, we have
\begin{eqnarray}
\label{FROMTHE6}
{{\sqrt{h}} \over N}\bigl(h^{mj}N^n-h^{nj}N^m\bigr)F^a_{0j}+N\sqrt{h}h^{mi}h^{nj}\epsilon_{ijk}B^k_a\nonumber\\
-{{\sqrt{h}} \over N}\bigl(h^{mi}N^nN^j+h^{nj}N^mN^i\bigr)\epsilon_{ijk}B^k_a=\beta\epsilon^{0mnj}F^a_{0j}.
\end{eqnarray}
\noindent
From the third term on the left hand side of (\ref{FROMTHE6}), we have the following relation upon relabelling indices $i\leftrightarrow{j}$ on the first term in brackets
\begin{eqnarray}
\label{FROMTHE7}
-h^{mi}N^nN^j\epsilon_{ijk}B^k_a-h^{nj}N^mN^i\epsilon_{ijk}B^k_a=-h^{mj}N^nN^i\epsilon_{jik}B^k_a-h^{nj}N^mN^i\epsilon_{ijk}B^k_a\nonumber\\
=\epsilon_{ijk}\bigl(h^{mj}N^n-h^{nj}N^m\bigr)N^iB^k_a.
\end{eqnarray}
\noindent
Note that the combination $h^{mj}N^n-h^{nj}N^m$ on the right hand side of (\ref{FROMTHE7}) is the same term multiplying $F^a_{0j}$ in the left hand side of (\ref{FROMTHE6}).  Using this fact, then (\ref{FROMTHE6}) can be written as
\begin{eqnarray}
\label{FROMTHE8}
{{\sqrt{h}} \over N}\Bigl[\bigl(h^{mj}N^n-h^{nj}N^m\bigr)\bigl(F^a_{0j}+\epsilon_{jki}B^k_aN^i\bigr)\Bigr]+\underline{N}\epsilon^{mnl}h_{lk}B^k_a=\beta\epsilon^{mnj}F^a_{0j}
\end{eqnarray}
\noindent
where $\epsilon^{0mnj}=\epsilon^{mnj}$.  Using $F^a_{0j}+\epsilon_{jki}B^k_aN^i=-\beta\underline{N}h_{jk}B^k_a$ from (\ref{FROMTHE3}) in (\ref{FROMTHE8}), then we have
\begin{eqnarray}
\label{FROMTHE9}
-{{\sqrt{h}} \over N}\bigl(h^{mj}N^n-h^{nj}N^m\bigr)\beta\underline{N}h_{jk}B^k_a+\underline{N}\epsilon^{mnl}h_{lk}B^k_a=\beta\epsilon^{mnj}F^a_{0j}.
\end{eqnarray}
\noindent
This simplifies to
\begin{eqnarray}
\label{FROMTHE10}
-beta\bigl(\delta^m_kN^n-\delta^n_kN^m\bigr)B^k_a+\underline{N}\epsilon^{mnl}h_{lk}B^k_a=\beta\epsilon^{mnj}F^a_{0j}\nonumber\\
\longrightarrow\beta\bigl(\epsilon^{mnj}F^a_{0j}+B^m_aN^n-B^n_aN^m\bigr)=\underline{N}\epsilon^{mnj}h_{jk}B^k_a.
\end{eqnarray}
\noindent
Contracting (\ref{FROMTHE10}) with $\epsilon_{mnl}$ and dividing by $2$, we obtain the relation
\begin{eqnarray}
\label{FROMTHE11}
F^a_{0l}+\epsilon_{lmn}B^m_aN^n-{1 \over \beta}\underline{N}h_{lk}B^k_a=0.
\end{eqnarray}
\noindent
Consistency of (\ref{FROMTHE11}) with (\ref{FROMTHE3}) implies that $\beta^2=-$, or that $\beta=\pm{i}$.

\section{Appendix B: Urbantke metric components}

\noindent
We now perform a 3+1 decomposition of the Urbantke metric\footnote{We have omitted the conformal factor for simplicity, which can always be re-inserted at the end of the derivations.}
\begin{eqnarray}
\label{URBAN}
g_{\mu\nu}=f_{abc}F^a_{\mu\rho}F^b_{\alpha\beta}F^c_{\sigma\nu}\epsilon^{\rho\alpha\beta\sigma}.
\end{eqnarray}
\noindent
In what follows we define $\epsilon^{0123}=1$, and make use of the fact that the structure constants $f_{abc}=\epsilon_{abc}$ for $SO(3,C)$ are numerically the same as the three dimensional epsilon symbol.  Also, we will use the definition $B^i_a={1 \over 2}\epsilon^{ijk}F^a_{jk}$ of the Ashtekar magnetic field.  The main result of this appendix is that due to the symmetries of the four dimensional epsilon tensor, each term in the expansion is the same to within a numerical constant.  We will show this by explicit calculation.\par
\noindent
(i) Starting from the time-time component we have
\begin{eqnarray}
\label{URBAN1}
g_{00}=f_{abc}f_{abc}F^a_{0\rho}F^b_{\alpha\beta}F^c_{\sigma{0}}\epsilon^{\rho\alpha\beta\sigma}.
\end{eqnarray}
\noindent
The time-time component of $g_{\mu\nu}$ reduces from two terms to one term, 
\begin{eqnarray}
\label{URBAN11}
f_{abc}F^a_{0i}F^b_{0j}F^c_{k0}\epsilon^{i0jk}+f_{abc}F^a_{0i}F^b_{j0}F^c_{k0}\epsilon^{ijk0}
=2f_{abc}\epsilon^{ijk}F^a_{0i}F^b_{0j}F^c_{0k}=12\hbox{det}(F^a_{0i}).
\end{eqnarray}
\noindent
(ii) Moving on to the space-time components, we have
\begin{eqnarray}
\label{URBAN2}
g_{0k}=f_{abc}F^a_{0\rho}F^b_{\alpha\beta}F^c_{\sigma{k}}\epsilon^{\rho\alpha\beta\sigma}
=f_{abc}F^a_{0i}F^b_{\alpha\beta}F^c_{\sigma{k}}\epsilon^{i\alpha\beta\sigma}\nonumber\\
=f_{abc}F^a_{0i}F^b_{0j}F^c_{lk}\epsilon^{i0jl}+f_{abc}F^a_{0i}F^b_{j0}F^c_{lk}\epsilon^{ij0l}+f_{abc}F^a_{0i}F^b_{jl}F^c_{0k}\epsilon^{ijl0}\nonumber\\
=-2\epsilon^{ijl}f_{abc}F^a_{0i}F^b_{0j}F^c_{lk}+f_{abc}F^a_{0i}F^c_{0k}B^i_b
=-f_{abc}F^a_{0i}F^b_{0j}B^m_c\epsilon^{ijl}\epsilon_{lkm}+f_{abc}F^a_{0i}F^c_{0k}B^i_b\nonumber\\
=-f_{abc}F^a_{0i}F^b_{0j}B^m_c\bigl(\delta^i_k\delta^j_m-\delta^i_m\delta^j_k\bigr)+f_{abc}F^a_{0i}F^c_{0k}B^i_b\nonumber\\
=f_{abc}\bigl(F^a_{0m}F^b_{0k}B^m_c-F^a_{0k}F^b_{0m}B^m_c\bigr)+f_{acb}F^a_{0m}F^b_{0k}B^m_c\nonumber\\
=-f_{abc}F^a_{0k}F^b_{0m}B^m_c=-(\hbox{det}F_{ok})\epsilon_{kml}(F^{-1})^{0l}_cB^m_c.
\end{eqnarray}
\noindent
(iii) The spatial components are given by
\begin{eqnarray}
\label{URBAN3}
g_{ij}=f_{abc}F^a_{i\rho}F^b_{\alpha\beta}F^c_{\sigma{j}}\epsilon^{\rho\alpha\beta\sigma}\nonumber\\
\end{eqnarray}
\noindent
which decomposes into a sum of four terms
\begin{eqnarray}
\label{URBAN4}
f_{abc}F^a_{i0}F^b_{mk}F^c_{lj}\epsilon^{0mkl}+f_{abc}F^a_{im}F^b_{0n}F^c_{ij}\epsilon^{m0nl}\nonumber\\
+f_{abc}F^a_{im}F^b_{n0}F^c_{lj}\epsilon^{mn0l}+f_{abc}F^a_{im}F^b_{nl}F^c_{0j}\epsilon^{mnl0}.
\end{eqnarray}
\noindent
Proceeding from (\ref{URBAN4}), we have
\begin{eqnarray}
\label{URBAN41}
f_{abc}F^a_{0i}B^l_bF^c_{lj}-f_{abc}\epsilon^{mnl}F^a_{im}F^b_{0n}F^c_{lj}
+f_{abc}F^a_{im}F^b_{n0}F^c_{lj}\epsilon^{mnl}-f_{abc}F^a_{im}B^m_bF^c_{0i}\nonumber\\
=f_{abc}F^a_{0i}B^l_bF^c_{lj}+f_{abc}F^a_{j0}B^l_bF^c_{li}+2f_{abc}\epsilon^{mnl}F^a_{im}F^b_{n0}F^c_{lj}.
\end{eqnarray}
\noindent
Note how the first two terms on the last line of (\ref{URBAN4}) are symmetric in $i$ and $j$.  Proceeding along, we have
\begin{eqnarray}
\label{URBAN5}
g_{ij}={1 \over 2}f_{abc}F^a_{i0}B^l_b\epsilon_{ljm}B^m_c+{1 \over 2}f_{abc}F^a_{j0}B^l_b\epsilon_{lim}B^m_c+2f_{abc}\epsilon^{mnl}F^a_{im}F^b_{n0}F^c_{lj}\nonumber\\
=f_{abc}F^a_{0i}f_{bcd}(B^{-1})^d_j(\hbox{det}B)+f_{abc}F^a_{0j}f_{bcd}(B^{-1})^d_i(\hbox{det}B)\nonumber\\
+{1 \over 2}f_{abc}\epsilon^{mnl}F^b_{n0}\epsilon_{imr}\epsilon_{ljs}B^r_aB^s_c.
\end{eqnarray}
\noindent
The result is to yield the two terms
\begin{eqnarray}
\label{URBAN6}
g_{ij}=2(\hbox{det}B)\bigl(F^a_{0i}(B^{-1})^a_j+F^a_{0j}(B^{-1})^a_i\bigr)+{1 \over 2}(\hbox{det}B)(B^{-1})^b_qF^b_{0n}\epsilon^{nlm}\epsilon^{qrs}\epsilon_{mri}\epsilon_{slj}\nonumber\\
\end{eqnarray}
\noindent
It can be shown by index manipulations that the quartic product of epsilon symbols reduces to
\begin{eqnarray}
\label{URBAN7}
\epsilon^{rsq}\epsilon^{mnl}\epsilon_{imr}\epsilon_{ljs}=\delta^q_i\delta^n_j+\delta^q_j\delta^n_i.
\end{eqnarray}
\noindent
Therefore, both contributions to (\ref{URBAN7}) are equal to within a numerical factor, and the spatial part of $g_{\mu\nu}$ is given by
\begin{eqnarray}
\label{URBAN8}
g_{ij}={5 \over 2}(\hbox{det}B)\bigl(F^a_{0i}(B^{-1})^a_j+F^a_{0j}(B^{-1})^a_i\bigr).
\end{eqnarray}

\end{document}